\documentclass[twocolumn]{aastex62}

\usepackage{amsmath,amssymb,amstext}
\usepackage{subfigure}

\usepackage{float}
\usepackage[version=4]{mhchem}
\usepackage{multirow}
\usepackage[utf8]{inputenc}
\usepackage{lmodern}

\usepackage{mathtools,nccmath}%

\usepackage{environ}

\usepackage{natbib}
\usepackage{hyperref}

\usepackage{xcolor}
\newcommand{\Mpc}{\,Mpc$^{-3}$}

\newcommand{\yr}{\,yr$^{-1}$}
\newcommand{\s}{\,s$^{-1}$}
\newcommand{\ergs}{\mbox{\,erg\,\s}}
\newcommand{\ergym}{\mbox{\,erg\,\Mpc\,\yr}}

\newcommand{\rj}{R_{\rm jet}}
\newcommand{\hj}{H_{\rm jet}}

\newcommand{\qi}{q}

\newcommand{\geff}{\Gamma_{\rm eff}}

\newcommand{\lbol}{L_{\rm bol}}
\newcommand{\lx}{L_{\rm x}}
\newcommand{\qcr}{Q_{\rm UHECR}}

\def\app#1#2{%
  \mathrel{%
    \setbox0=\hbox{$#1\sim$}%
    \setbox2=\hbox{%
      \rlap{\hbox{$#1\propto$}}%
      \lower1.1\ht0\box0%
    }%
    \raise0.25\ht2\box2%
  }%
}
\def\approxprop{\mathpalette\app\relax}

\begin{document}

\title{High-Energy Neutrino Emission from Espresso-Reaccelerated Ions in Jets of Active Galactic Nuclei}

\author[0000-0001-9475-5292]{Rostom Mbarek}
\email{rmbarek@uchicago.edu}
\affiliation{University of Chicago, Department of Astronomy \& Astrophysics, 5640 S Ellis Ave., Chicago, IL 60637, USA}
\affiliation{Kavli Institute for Cosmological Physics, The University of Chicago, Chicago, IL 60637, USA}
\affiliation{Enrico Fermi Institute, The University of Chicago, Chicago, IL 60637, USA}

\author[0000-0003-0939-8775]{Damiano Caprioli}
\affiliation{University of Chicago, Department of Astronomy \& Astrophysics, 5640 S Ellis Ave., Chicago, IL 60637, USA}
\affiliation{Enrico Fermi Institute, The University of Chicago, Chicago, IL 60637, USA}

\author[0000-0003-0939-8775]{Kohta Murase}
\affiliation{Department of Physics, The Pennsylvania State University, University Park, Pennsylvania 16802, USA}
\affiliation{Department of Astronomy \& Astrophysics, The Pennsylvania State University, University Park, Pennsylvania 16802, USA}
\affiliation{Center for Multimessenger Astrophysics, Institute for Gravitation and the Cosmos, The Pennsylvania State University, University Park,
Pennsylvania 16802, USA}
\affiliation{Center for Gravitational Physics, Yukawa Institute for Theoretical Physics, Kyoto, Kyoto 606-8502 Japan}

\begin{abstract}
We present a bottom-up calculation of the flux of ultra-high energy cosmic rays (UHECRs) and high-energy neutrinos produced by powerful jets of active galactic nuclei (AGNs). By propagating test particles in 3D relativistic magnetohydrodynamic jet simulations, including a Monte Carlo treatment of sub-grid pitch-angle scattering and attenuation losses due to realistic photon fields, we study the spectrum and composition of the accelerated UHECRs and estimate the amount of neutrinos produced in such sources.
We find that UHECRs may not be significantly affected by photodisintegration in AGN jets where the \emph{espresso} mechanism efficiently accelerates particles, consistent with Auger's results that favor a heavy composition at the highest energies.
Moreover, we present estimates and \emph{upper bounds} for the flux of high-energy neutrinos expected from AGN jets.
In particular, we find that: i) source neutrinos may account for a sizable fraction, or even dominate, the expected flux of cosmogenic neutrinos; 
ii) neutrinos from the $\beta$-decay of secondary neutrons produced in nucleus photodisintegration end up in the TeV--PeV band observed by IceCube, but can hardly account for the observed flux;
%iii) since the most important background for UHECR--photons interactions is the AGN non-thermal emission, a picture arises where high-energy neutrino emission could correlate with AGN flaring activity.
iii) UHECRs accelerated via the \emph{espresso} mechanism lead to nearly isotropic neutrino emission, which suggests that nearby radio galaxies may be more promising as potential sources.
We discuss our results in the light of multimessenger astronomy and current/future neutrino experiments.
\end{abstract}

\defcitealias{murase+14}{MID14}
\defcitealias{mbarek+19}{MC19}
\defcitealias{mbarek+21a}{MC21}

\section{Introduction}
A comprehensive theory that accounts for particle injection, particle acceleration, and spectra of ultra-high energy cosmic rays (UHECRs) with energies above $10^{18}$eV is still lacking. 
The origin of these particles, however, has been the subject of many theoretical studies that rely mostly on estimates of the maximum energy that particles can achieve in specific environments \citep{cavallo78,hillas84,UHECRrev19}. 
In that regard, one of the most promising astrophysical sources of UHECRs are AGN jets,
which satisfy the Hillas criterion up to $10^{20}$eV, especially if the highest-energy CRs are heavy ions. 
Additionally, their luminosities can explain the energy injection rate necessary to sustain the UHECR flux \citep[e.g.,][]{katz+09,murase+19,jiang+21}. 
Other sources have also been suggested such as newly-born millisecond pulsars \citep[e.g.,][]{fang+12,fang+14}, 
$\gamma$-ray bursts \citep[e.g.,][]{vietri95,waxman95}, engine-driven transrelativistic supernovae including low-luminosity $\gamma$-ray bursts and hypernovae \citep[e.g.,][]{murase+06,wang+08,zm19}, and tidal disruption events \citep[e.g.,][]{farrar+14,zhang+17}. However, the exact acceleration mechanism boosting UHECRs to their energies in these environments is not pinned down.

This paper is the third in a series of projects aimed to analyze the promising \emph{espresso} model \citep{caprioli15}, checking if it may satisfy all the requirements of a particle acceleration theory and outlining its observational predictions.
In essence, the \emph{espresso} framework suggests that UHECRs can be produced in relativistic AGN jets via the reacceleration of galactic CR seeds.
Such seeds, accelerated in supernova remnants up to a few PV in rigidity, penetrate in the highly relativistic regions of the jets and tap in their radial electric field to receive one, or even multiple, $\sim \Gamma^2$ boosts in energy. 
If the jet is sufficiently powerful with $\Gamma \sim 20-30$, a single shot would allow them to reach UHECR energies. 
In \citet{mbarek+19,mbarek+21a}, the \emph{espresso} mechanism has been tested by propagating particles in high-resolution magnetohydrodynamical (MHD) simulations of AGN jets \citep{mignone+07,rossi+08}. 
The reacceleration of galactic CR seeds is also promising even for subrelativistic AGN jets that are seen at kpc scales, and the model can fit the Auger spectrum and composition data simultaneously \citep{kimura+18}.

In \citet{mbarek+19}, hereafter \citetalias{mbarek+19}, we found that the spectra, chemical composition, and anisotropy of the reaccelerated particles are qualitatively consistent with UHECR phenomenology. 
Then, in \citet{mbarek+21a}, hereafter \citetalias{mbarek+21a}, we included sub-grid scattering (SGS) to model small-scale magnetic turbulence that cannot be resolved by MHD simulations, constraining for the first time one potentially crucial but hard-to-model ingredient. 
We established the relative importance of \emph{espresso} and stochastic shear acceleration in relativistic jets, finding that strong SGS, on one hand, can promote the injection and acceleration of lower-energy UHECRs, but on the other hand, is irrelevant for the acceleration of the highest-energy CRs, which are invariably \emph{espresso}-accelerated in the inner regions of relativistic jets. However, we should also keep in mind that shear acceleration may play a more important role in accelerating particles in subrelativistic jets that are usually seen in kiloparsec scales especially for Fanaroff-Riley (FR) I galaxies. 
Overall, we expect the neutrino production to be relatively important in the relativisitic spine of the jet, in particular powerful jets seen in FR II galaxies and flat spectrum radio  quasars (FSRQs), where particles are promoted to the highest energies.

In this paper, we evaluate the effects of photodisintegration and high-energy neutrino production in AGN jets within our self-consistent particle acceleration framework. 
In particular, we investigate how the intense radiation fields of the blazar zone, the broad-line region, and the dusty torus may affect the chemical composition of the accelerated particles.
Moreover, modeling UHECR attenuation in a realistic jet environment allows us to calculate the spectrum of high-energy neutrinos produced in these sources.

%In this regard, this paper is the last cornerstone in rendering \emph{espresso} a holistic particle acceleration framework. 

Within our bottom-up approach we aim to address, with as few assumptions as possible, some key open questions such as:
\begin{itemize}
    \item What are the effects of losses on \emph{espresso}-accelerated particles in sub-kiloparsec-scale AGN jets?
    \item What is the expected spectrum of UHE (Ultra-High-Energy) neutrinos produced by a typical AGN jet?
    \item If UHECRs are accelerated in AGN jets via the \emph{espresso} mechanism, can they be responsible for the observed IceCube flux, too?
\end{itemize}

This work would be particularly important to unravel questions associated with Ultra-High-Energy (UHE) neutrinos.
UHE neutrinos are created through interactions of UHECRs and are pivotal tools to advance our knowledge of extreme astrophysical environments. 
Many current and proposed experiments, such as the balloon-borne interferometer ANITA \citep{Anita18a-long,Anita18b-long}, its successor the Payload for Ultrahigh Energy Observations (PUEO) \citep{PUEO21}, the Askaryan Radio Array (ARA) \citep{ARA12-long,ARA16-long}, the In-Ice Radio Array ARIANNA \citep{Arianna17-long}, the exciting next-generation IceCube detector IceCube-Gen2 \citep{ICECUBE-GEN2-14}, the Giant Radio Array for Neutrino Detection (GRAND) \citep{GRAND20}, the Radio Neutrino Observatory in Greenland (RNO-G) \citep{RNO-G21}, the Beamforming Elevated Array for Cosmic Neutrinos (BEACON) \citep{wissel+20}, and the proposed POEMMA mission \citep{olinto+17}, all aim to detect EeV neutrinos for the first time. 
There are many studies that focus on \emph{cosmogenic neutrinos}, i.e., the neutrinos created by UHECRs through interactions with the extragalactic photon background during intergalactic propagation \citep[e.g.][]{beresinsky+69,yoshida+93,takami+09,heinze+16,romero-Wolf+18,heinze+19,das+19,wittkowski+19}. 
However, UHE \emph{source neutrinos}, i.e., neutrinos produced in or around UHECR accelerators, especially in the presence of extreme photon fields, could also be crucial to unravel the sites of production of the highest-energy particles in the Universe~\citep[see][and references therein]{batista+19}.
This study aims to shed more light on these UHE \emph{source neutrinos}.

%\textbf{Different studies have considered neutrino production in sources and set bounds on the neutrino diffuse flux \citep{waxman+98,mannheim+00}. \citet{waxman+98} derived a limit on the neutrino flux under the assumptions that the sources are optically thin and accelerated with a Fermi acceleration spectrum $\sim E^{-2}$ and that the proton energy is entirely transferred to pions. The \citet{mannheim+00} bound, on the other hand, exploits experimental results to infer an injection spectrum and consider thick sources. These landmark studies are considered quite general even if they only rely on nominal scales \citep{murase+10}. Other studies have calculated the expected neutrino events or performed correlation studies at the highest neutrino energies with specific sources classes such as supernovae \citep[e.g.,][]{pitik+22} $\gamma$-ray bursts \citep[e.g.,][]{petropoulou+14,meszaros17}, clusters of galaxies \citep[e.g.,][]{zandanel+15}, star-forming galaxies \citep[e.g.,][]{tamborra+14}, tidal disruption events \citep[e.g.,][]{wang+16,murase+20}, or the magnetized coronae of AGNs \citep[e.g.,][]{kheirandish+21}. Generally speaking, neutrino production from these sources relies on assumptions on initial particle acceleration where particles are accelerated through shock acceleration \citep[e.g.,][]{pitik+22} or magnetic reconnection \citep[e.g.,][]{khiali+15}.
In this paper, we propagate particles in an MHD simulation of an AGN jet to keep assumptions to a minimum and find that particles are \emph{espresso}-accelerated \citepalias{mbarek+19,mbarek+21a}. The neutrino spectrum that ensues is presented in \S\ref{sec:uhecrNus}.

The paper is organized as follows. 
In \S\ref{particle_propagation} we describe our particle acceleration framework, detailing the different interaction routes that lead to losses and neutrino production. In \S\ref{sec:uhecrNus} we investigate the effects of losses on the UHECR chemical composition and put constraints on the expected \emph{upper bounds} of the neutrino spectrum resulting from UHECR interactions.
We also discuss the acceleration mechanism responsible for boosting UHECRs that may contribute to the IceCube flux, and finally present our conclusions in \S\ref{sec:conc}.

\section{Propagating Particles in Realistic Jets}\label{particle_propagation}
\subsection{MHD simulation of a relativistic jet}
To facilitate the comparison with published results, we model the underlying AGN relativistic jet via the same benchmark simulation used in \citetalias{mbarek+19} and \citetalias{mbarek+21a}; we refer to those papers for all the details and summarize here the essential information.
We consider a 3D relativistic MHD simulation of a powerful AGN jet performed with PLUTO \citep{mignone+10}, which includes adaptive mesh refinement. 
The jet, with a magnetization radius $\rj$, is initialized with Lorentz factor $\Gamma_0 = 7$ along the $z$-direction in a box that measures 48$\rj$ in the $x$- and $y$-directions and 100$\rj$ in the $z$-direction in a grid that has $512 \times 512 \times 1024$ cells with four refinement levels. 
The jet/ambient density contrast is set to $\psi = 10^{-3}$, the jet sonic and Alfv\'enic Mach numbers to $M_s = $3, and $M_A = 1.67$ respectively.
Once the jet has developed, the effective Lorentz factor in the jet spine is $\Gamma_{\rm eff}\sim 3.2$; this value is important to establish how many $\Gamma^2$ shots a particle undergoes during acceleration.

\subsection{Particle propagation}
We propagate $\sim 10^5$ test particles in a snapshot of the benchmark jet with a broad range of initial gyroradii $\mathcal R$ and positions. 
We include the effects of unresolved turbulence by setting the SGS mean free path to be as small as the particle's gyroradius (Bohm diffusion) to maximize the number of particles within the jet spine and boost the efficiency of particle acceleration (see \citetalias{mbarek+21a} for more details on the effects of SGS). 
This prescription maximizes the particles residence time in the jet (and hence the effects of photon fields), but does not affect particle acceleration at the highest energies, which are invariably accelerated via the SGS-independent \emph{espresso} mechanism (see \citetalias{mbarek+21a}).

In addition to protons, we consider four different seed ion species, labelled $se=$[He, C/N/O,Mg/Al/Si, Fe] with effective atomic number $Z_{\rm se}=[2,7,13,26]$ and mass $A_{\rm se}= [4,14,27,56]$, respectively. 
The energy flux of these seed galactic CRs below the knee is parameterized as follows:
\begin{equation} \label{phi_eq}
\phi_{\rm se}(E) = K_{\rm se} \left( \frac{E}{10^{12}~\rm eV}\right) ^{-q_{\rm se}} ,
\end{equation}
We set the normalizations according to the abundance ratios at $10^{12}$~eV observed in Galactic CRs, such that $K_{\rm se}/K_H \sim$ [0.46, 0.30, 0.07,0.14]. 
The motivation for using Galactic CR fluxes as fiducial cases hinges on the fact that the knee feature (the maximum seed energy) should be quite independent of the galaxy mass \citep{caprioli15}.
We acknowledge that the actual seed fluxes in AGN hosts may be different due to the different injection and confinement properties of other galaxies but, since the chemical enrichment provided by diffusive shock acceleration of seeds in supernova remnants \citep{caprioli+10a,caprioli+17} is rather universal and since the final UHECR fluxes that we consider are normalized to the observed one, the assumptions above are quite generic.
The spectra of different species are calculated as outlined in Appendix~\ref{z_appendix}.

Particles are propagated in the MHD simulation even if their gyroradii are smaller than the grid size, which is reasonable as long as a sufficiently small rigidity-dependent timestep is used to resolve their gyration; every particle's gyroradius is resolved with  \emph{at least} 10 timesteps.

\subsection{Photon field prescriptions}
\begin{figure}
\centering
\includegraphics[width=0.5\textwidth,clip=false, trim= 0 0 0 0]{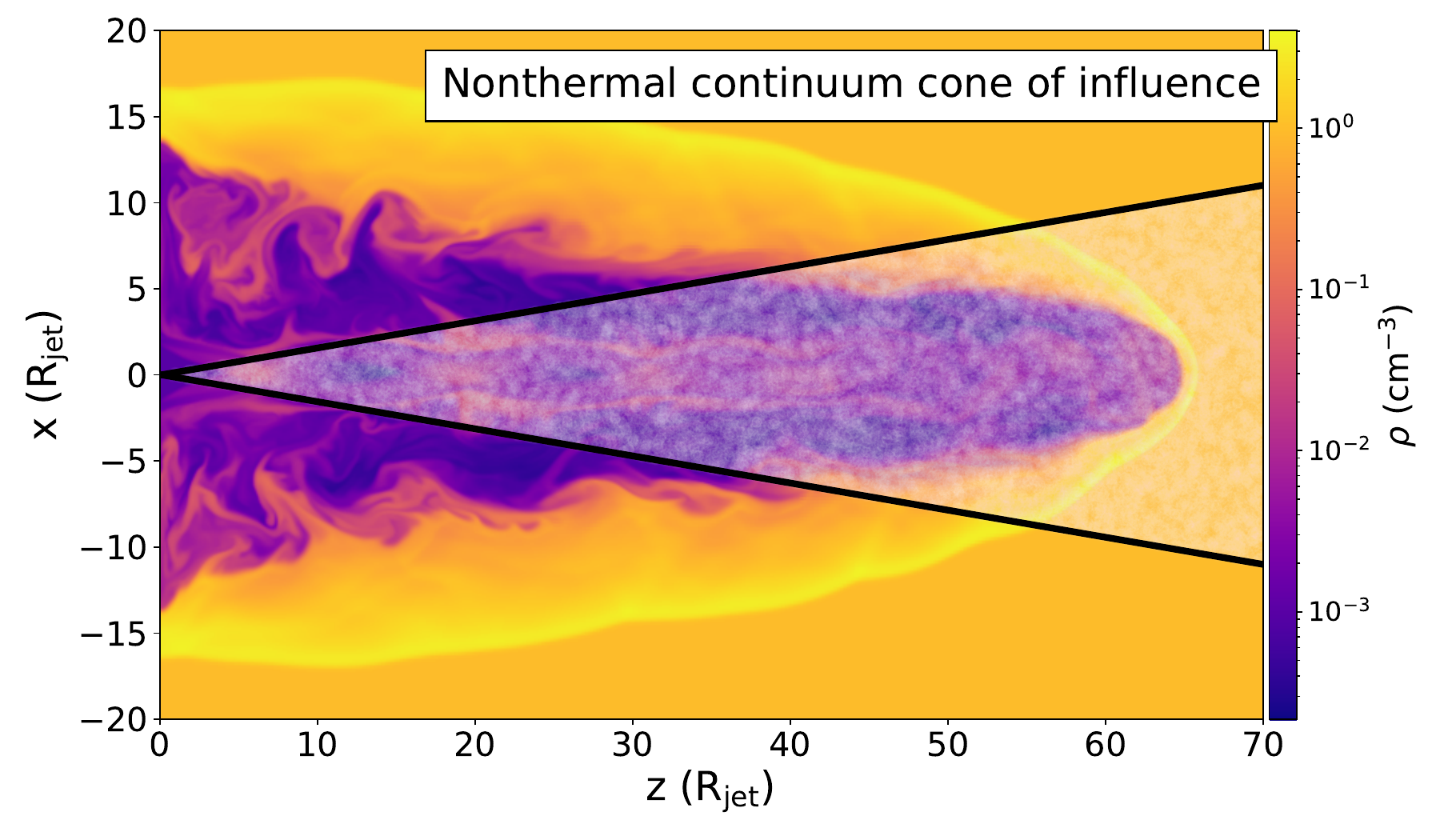}
\caption{Non-thermal continuum cone of influence overlaid on a 2D slice of the density component of the MHD jet. The cone is relativistically beamed at an angle $1/\geff$, Where $\geff \sim 3.2$ is the effective Lorentz factor of the jet (See Figure~6 in \citetalias{mbarek+19} for more details).
The density is normalized based on the assumptions discussed in \S\ref{pp-paragraph} to maximize $pp$ interactions.}
\label{cone}
\end{figure}

On top of the jet structure provided by the MHD simulation, we prescribe external photon fields based on the methods presented in \citet{murase+14}, hereafter \citetalias{murase+14}. There is ostensible uncertainty in modeling these external components due to the vast AGN diversity, but the systematic approach of \citetalias{murase+14} allows us to assess the individual effect of such fields based on the apparent bolometric luminosity of the jet.

We consider five different photon backgrounds of different origin: 
\paragraph{(i) Non-thermal emission:} 
It originates from  synchrotron and inverse-Compton radiation of relativistic leptons and/or hadronic emission, emerging from the blazar zone, a sub-pc region close to the base of the jet (see \citetalias{murase+14} for more details), where the emission is dominated by X/$\gamma$-rays and is relativistically beamed  with an angle $\sim 1/\Gamma$ in the black hole frame. 
We refer to this broadband emission region as the non-thermal cone of influence (see Figure~\ref{cone}). 
\paragraph{(ii) Radiation from the broad atomic line region (BLR):}
This is the reprocessed emission from cold gas clumps photoionized by the UV and X-ray produced by the accretion disk. 
These sub-pc spherical clumps are located closer to the base of the jet ($<1$pc away) and have a luminosity $\sim 10$\% of that of the accretion disk (see \citetalias{murase+14}).
\paragraph{(iii) IR emission from the dusty torus:}
This is IR from reprocessed accretion dusty disk radiation with a torus size that can reach $\sim 1$pc. 
Following \citetalias{murase+14}, we model it as a spherical grey body with temperature $\sim$500K.
\paragraph{(iv) Stellar light:} 
The photons from the host-galaxy stars have been shown to have large energy densities compared to other photon fields at a few hundred pc, which makes them important targets for accelerated particles in powerful AGN jets. 
In the remainder of the paper, we will consider the starlight emission profile for Centaurus A as our fiducial case \citep{tanada+19}.

\paragraph{(v) The cosmic microwave background (CMB) radiation:} Besides affecting every particle regardless of its location, accounting for the CMB contribution serves as a benchmark to compare the effect of the prescribed photon fields.
 
In the remainder of the paper, all photon fields are assumed to be isotropic except for the non-thermal component, which is beamed within a cone of aperture $1/\geff$, where $\geff$ is the effective Lorentz factor of the jet. 
The BLR, IR and non-thermal contributions have intensities that are inversely proportional to the square of the distance from their emitting regions.

\begin{figure}
\centering
\includegraphics[width=0.49\textwidth,clip=false, trim= 0 0 0 0]{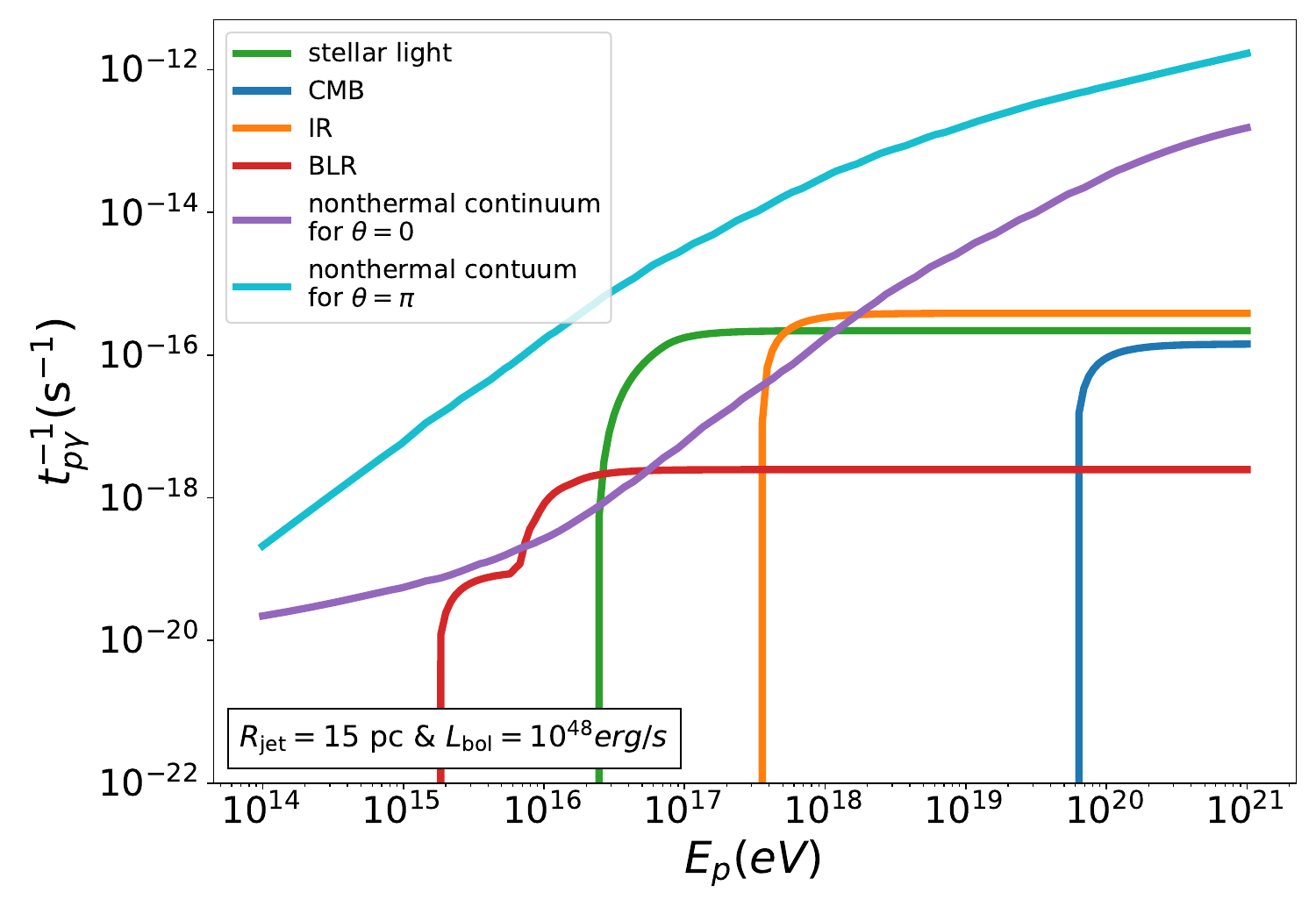}
\includegraphics[width=0.49\textwidth,clip=false, trim= 0 0 0 0]{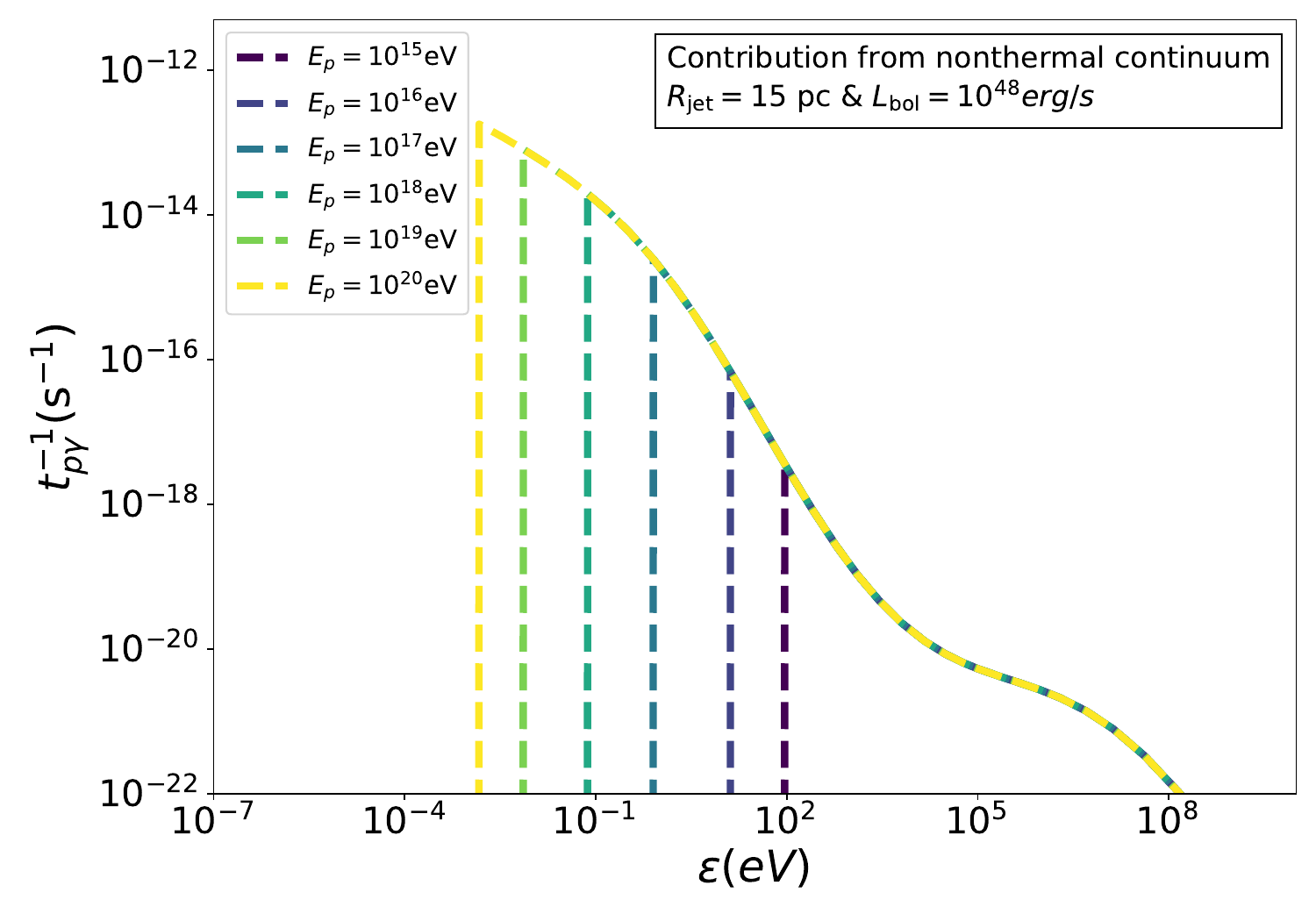}
\caption{Upper Panel: Photomeson cooling time for a proton located at a position (0,0,750pc) for a jet with $\rj = 15$pc and $\lbol = 10^{48}\ergs$, where $\lbol$ is the isotropic-equivalent bolometric luminosity.
Contributions from the isotropic photon fields (BLR, IR, stellar light, and CMB) are calculated based on equation~\ref{s3}. Contributions from the beamed non-thermal continuum are calculated based on equation~\ref{pg-angle}. Here, $\theta$ denotes the angle between the momentum of the proton and the target photon such that $\theta = \pi$ for head-on interactions and $\theta = 0$ for tail-on interactions. Lower Panel: Photomeson cooling time for protons with different energies located at the same position as a function of the photon energy $\epsilon$. This plot only considers head-on interactions ($\theta = \pi$) with the non-thermal continuum as an example.}
\label{tpg}
\end{figure}

\subsection{Particle interactions}
\begin{table*}
\begin{centering}
\begin{tabular}{ |c|c|c|c|c| } 
\hline
Particle & Process & Reactions & Neutrino Energy fraction \\
\hline
\hline
\multirow{2}{*}{Proton (p)} & proton-proton ($pp$) & $\ce{p}+\ce{p} \rightarrow \ce{p} + \ce{n} + \pi^{\text{+}} \rightarrow \ce{p} + \ce{n}+e^{\text{+}}+\nu_e+\nu_\mu+ \bar{\nu}_\mu$ & p:$\nu \sim $ 20:1 \\
& photomeson ($p\gamma$) & $\ce{p}+\gamma \rightarrow \ce{n} +\pi^{\text{+}} \rightarrow \ce{n}+e^{\text{+}}+\nu_e+\nu_\mu+ \bar{\nu}_\mu$ & p:$\nu \sim $ 20:1 \\
\hline
\multirow{2}{*}{Nucleus (N)} & photomeson ($N\gamma$) & $\ce{N}+\gamma \rightarrow \ce{^{A-1}N}+\ce{n}+ \pi^{\text{+}} \rightarrow \ce{^{A-1}N}+\ce{n}+ e^{\text{+}}+\nu_e+\nu_\mu+ \bar{\nu}_\mu$ & N:$\nu \sim $ 20 A:1 \\ 
& photodisintegration & $\ce{^{A}N}+\gamma \rightarrow \ce{^{A-1}N}+ \ce{n}\rightarrow \ce{^{A-1}N}+ \ce{p}+e^-+\bar{\nu}_e$ & N:$\nu \sim 2\times 10^3 $A:1 \\
& \& neutron decay & $\ce{^{A}N}+\gamma \rightarrow \ce{^{A-1}N}+ \ce{p}$ & --- \\

\hline
\end{tabular}
\caption{\label{tab:interactions} Neutrino production mechanisms for protons ($p$) and nuclei ($N$) of atomic mass $A$.
The last column gives the ratio of the energy of the neutrino with respect to the parent particle (the parameter $\alpha$ is introduced in Appendix~\ref{sp_appendix}).}
\end{centering}
\end{table*}

When propagating our test particles, at every time step we: (i) calculate the probability of interaction with the thermal plasma (assuming it is electron--proton) and photon fields; 
(ii) keep track of each particle's atomic mass A and charge Z; 
(iii) monitor secondary particle production including neutrinos and secondary protons. Finally, secondary protons and ions are further propagated until parent and secondary nuclei have travelled a distance of  at least 100 $\rj$. 
Produced neutrinos are assumed to escape without  experiencing further interactions; 
their place of production and escaping direction are recorded, though.

In order to study the effects of UHECR photodisintegration and the resulting neutrino flux, we consider the most relevant attenuation mechanisms, i.e., inelastic proton-proton ($pp$), photomeson production ($p\gamma$), and neutron decay following photodisintegration of heavy nuclei. 
More specifically, photomeson production interactions of photons and nucleons also result in the production of pions that subsequently decay to create photons and neutrinos. 
On the other hand, photodisintegration interactions are nuclear processes that cause the photon-absorbing nucleus to change to another chemical specie and release either a proton or a neutron.
Table~\ref{tab:interactions} summarizes the interaction routes that lead to neutrinos and photodisintegration of heavy elements. 
For simplicity, we ignore the Bethe-Heitler process in this work, which is sufficient for our current purpose. This process can be important for setting the maximum energy only for very powerful blazars \citepalias{murase+14}. 
A more detailed explanation of the interaction probabilities at every time step is included below.

\subsubsection{Proton-proton ($pp$) interactions}\label{pp-paragraph}

Accelerated particles can experience $pp$ scattering to create charged pions and hence $\nu_e$ and $\nu_\mu$ neutrinos. At every time step, depending on the particle's energy $E_p$ and position $\mathbf{x}$, there is an interaction probability $P(E_p, \mathbf{x}) \sim  n(\mathbf{x}) \sigma_{pp}(E_p) R_{\rm jet} \Delta t$, where $\sigma_{pp}$ is the $pp$ interaction cross section \citep{PDG18}, $n$ is the position-dependent density, and $\Delta t$ is the time step. The neutrino spectrum that results from every $pp$ interaction is calculated based on the parametrization by \citet{kelner+06}.

Powerful AGN jets are often inside clusters of galaxies \citep[e.g.,][]{begelman+84,best+07,fang+18}, so that the ambient medium density should reflect that of the intra-cluster medium, which is of order of $n_{\rm ICM} \sim 10^{-3}$cm$^{-3}$ \citep{walg+13}. This reference value is used in Figure~\ref{cone}, which also shows that the jet itself is expected to have an even lower density. 
Generally, $pp$ interactions inside the AGN jets are not expected to contribute much to the overall neutrino spectrum, although they can be relevant for cosmic rays that have escaped from AGN jets and are confined in clusters \citep{fang+18}.

\subsubsection{Photomeson production ($p\gamma$) Interactions} \label{pg-inter}
At every time step $\Delta t$, a photopion production probability $f_{p\gamma}$ is calculated such that $f_{p\gamma} = t_{p\gamma}^{-1} \Delta t $, where $t_{p\gamma}$ is the photomeson cooling time. 
A detailed account of $t_{p\gamma}$ calculations for isotropic photon fields and interactions at a known angle for both protons and nuclei with atomic mass A is provided in Appendix~\ref{photomes_appendix}. For $p\gamma$ interactions, we assume a multiplicity of 1 for neutrinos (with energies $E_\nu \sim E_p/20$) produced in one interaction of a relativistic proton, which is valid when the resonance and direct production are relevant. This assumption is valid when target photon spectra are sufficiently soft \citep{murase+08a}. 

In general, the cooling time depends on the particle position and on the AGN luminosity; therefore, we cannot use dimensionless quantities but need to introduce physical scales for the magnetic field strength and for the jet size and luminosity. 
For instance, the left panel of Figure~\ref{tpg} shows the cooling time for protons of different energies located 750 pc away from the base of the jet along the spine ($z$-direction in Figure~\ref{cone}), for a jet with radius $\rj = 15$pc and an apparent bolometric luminosity $L_{\rm bol} = 10^{48}\ergs$. 
In this characteristic example, we show the contribution from isotropic photon fields (BLR, IR, stellar light, and CMB, as in the legend) and the angle-dependent interactions with the beamed non-thermal continuum (cyan and purple lines).  
The left panel of Figure~\ref{tpg} shows that the non-thermal contribution provides the shortest photomeson production cooling time (even for tail-on interactions, i.e., $\theta = 0$). 
These cooling curves depend on the particle position (distance from the base of the jet and position with respect to the non-thermal continuum cone of influence) as will be further discussed below. 

The right panel of Figure~\ref{tpg} shows the dependence of the cooling time on the photon energy $\epsilon$ when the proton energy is fixed. 
Only the non-thermal contribution is shown here for simplicity as it is the dominant photon field in our fiducial cases.
We can see that the shortest cooling time---which depends on the proton energy--- occurs at the threshold energy $\bar{\epsilon}_{\rm th}$ for photomeson production interactions (See Appendix \ref{photomes_appendix} for more details). 
Importantly, the contribution of X/$\gamma$-ray photons with $\epsilon \geq 100 $eV to cooling is only significant for lower energy protons.

\subsubsection{Photodisintegration interactions}\label{sec:phot-inter}
\begin{figure}
\begin{centering}
\includegraphics[width=0.48\textwidth,clip=false,trim= 0 0 0 0]{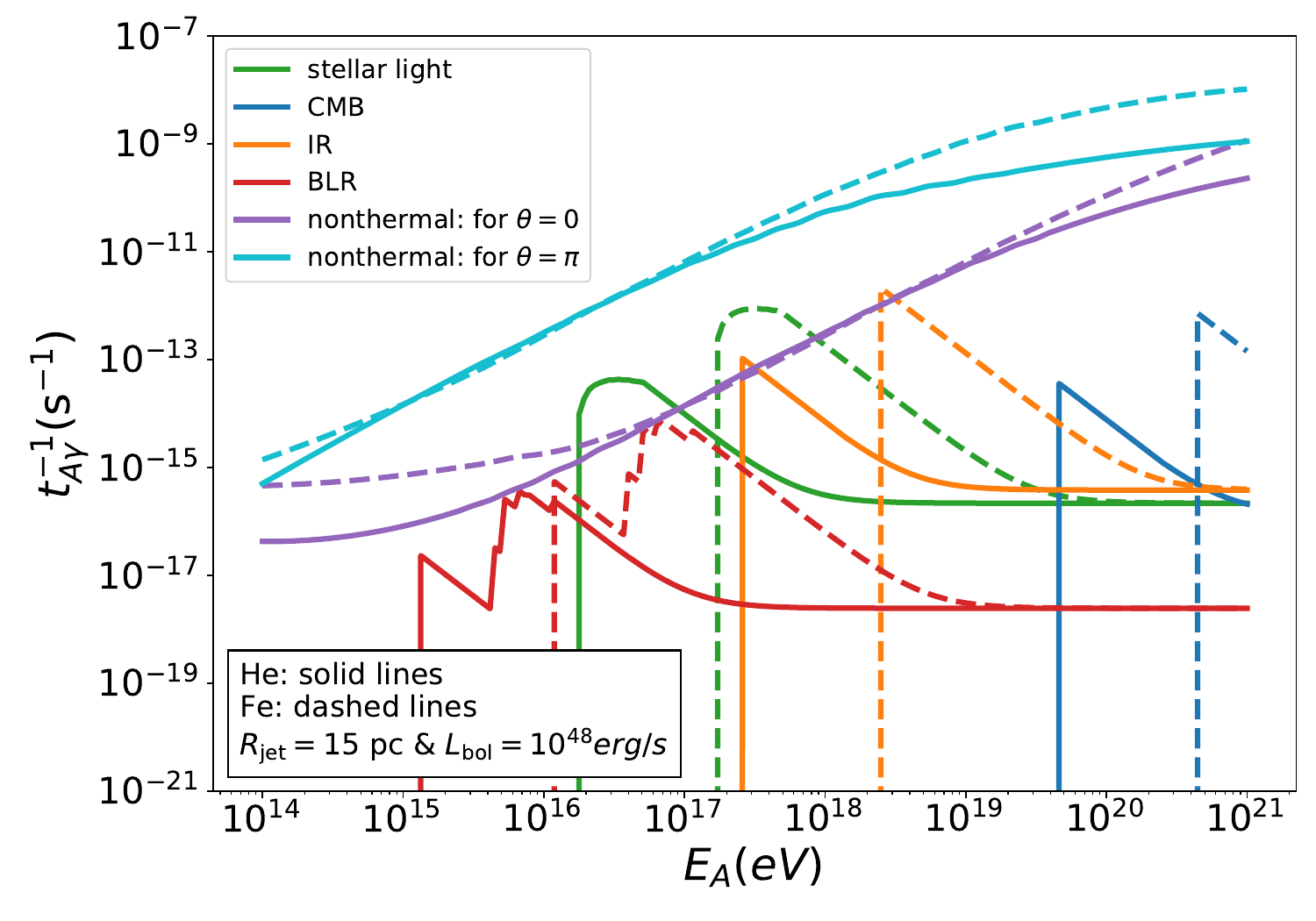}
\caption{Same as the Upper panel of Figure~\ref{tpg} but for photodisintegration interactions of nuclei with energy $E_A$, including photomeson interactions and interactions based on the GDR photodisintegration total cross section (see Equation~\ref{eq:phot} and~\ref{eq:phot-angle}).  
Note that the threshold for He photodisintegration is quite close to that of photomeson production because the plot is a function of $E_A$ and a factor of $A$ must be taken into consideration.
}
\label{tag}
\end{centering}
\end{figure}

On the same photon fields, nuclei with atomic mass $A$ can also undergo photodisintegration with probability $t_{A\gamma}^{-1} \Delta t $ per time step.
Appendix~\ref{photodis_appendix} details our calculations, in particular the use of the giant dipole resonance (GDR) cross section \citep[e.g.,][]{murase+08a,wang+08} as a fiducial case (see Figure~\ref{tag}). Although this is the simplest application, it is sufficient to demonstrate the effects of photodisintegration in our numerical work. The GDR approximation is valid for soft photon spectra \citep{murase+08a}, and non-thermal photon fields are dominant in our examples. More detailed studies including quasi-deutron emission and fragmentation are left as future work. 
We also account for photomeson production interactions of heavy nuclei (as described in Equation~\ref{s3n} in Appendix~\ref{photomes_appendix}), and find that their cooling time is comparable to that of protons.
Also for nuclei, the most important photon background is typically the non-thermal component, as shown by a comparison of Figure~\ref{tag} with Figure~\ref{tpg} and Equation~\ref{s3n}. Higher-energy nuclei are more likely to be photodisintegrated, as expected. Note that the thresholds for photodisintegration interactions for He look quite close to those of photomeson interactions because they are plotted as a function of $E_A$, hence a scaling by the atomic mass A is necessary.  

While propagating particles, we keep track of the atomic mass and charge of nuclei, considering that after a photodisintegration event, the nucleus loses either one neutron or one proton.
A neutron produced as a result of photodisintegration decays within a distance 9.15($E_n/10^9$GeV)~kpc \citep{anchordoqui+07} to produce one neutrino with energy 0.48 MeV in the neutron rest frame \citep{murase+10} such that $E_n/E_\nu \approx 2\times 10^3$ in the lab frame.
Since secondary particles cannot achieve energies beyond $\sim 10^{10}$GeV, we assume that all the secondary neutrons $\beta$-decay and produce neutrinos.

\subsubsection{Dependence on the distance from the central black hole}\label{sec:d-dependence}
Since different photon backgrounds have different spatial extents, the cooling time for both photodisintegration and $p\gamma$ interactions depend on the magnitude of the distance from the base of the jet, $D$, such that close to the base of the jet $t^{-1} \propto D^{-2}$ where photon fields other than the CMB are most relevant.
For a jet with isotropic-equivalent bolometric luminosity $\lbol = 10^{45}\ergs$, non-thermal emission is dominant until $D\sim 3$kpc, beyond which the CMB becomes important. Considering that the photon fields we assume are generated close to the base of the jet, CR interactions beyond a few kpc would not increase with large jet extents, usually associated with the more luminous FR-II jets.

\subsection{From scale-free MHD simulations to realistic environments}\label{jet-lum}
In our simulations, CR gyroradii are normalized to the jet radius $\rj$ and magnetic field $B_0$; therefore, setting a physical value to $\rj$ and $B_0$ is equivalent to associating physical energies to the seed particles of charge $Ze$. 
In order to calculate the actual neutrino fluxes from realistic AGNs, we need to fix a reference magnetic field and two physical quantities of the jet: its bolometric luminosity, which controls the photon fields, and its radius.

Our simulations have extensively shown that particles are routinely \emph{espresso}-accelerated up to the jet Hillas limit (\citetalias{mbarek+19}, \citetalias{mbarek+21a}).
In our 3D relativistic MHD simulations, we launch the jet with $\Gamma=7$; 
yet, the effective Lorentz factor of evolved jets turns out to be $\geff\sim$ a few, too small to promote CR seeds with rigidities of a few PeV to actual UHECRs with a single $\geff^2$ boost. Therefore, in order to achieve realistic UHECR energies, we fix the normalization of our jet radius and magnetization such that CR seeds have rigidities as large as $3\times 10^{17}$V, two orders of magnitude above the CR knee.
This choice allows us to include the attenuation losses discussed for realistic photon fields and to calculate the fluxes of UHECRs and HE neutrinos expected from different types of AGNs.

When contemplating assigning magnetic field, radius, and $\lbol$ prescriptions to our jet, one needs to consider different types of radio-loud AGNs, which at minimum can be split into FR-I and FR-II sources. 
FR-I jets are typically decelerated to nonrelativistic bulk flows within 1 kpc \citep[e.g.,][]{wardle+97,arshakian+04,mullin+09}, while FR-II jets, show $\Gamma\gtrsim 10$ at scales of tens of kpc and beyond \citep[e.g.,][]{sambruna+02,siemiginowska+02,tavecchio+04,harris+06}.
The FR dichotomy likely reflects a combination of jet power and ambient density \citep{bromberg+16}, and our fiducial jet propagating in a homogeneous density profile may resemble an FR-I jet more than an FR-II one, with a %relatively low $\geff$ and a 
small $\rj/\hj$ ratio, where $\hj$ is the extent of the jet.
Yet, in this work we consider a broad range of luminosities that should span the higher-luminosity parameter space.
More precisely, we consider the two following cases.

\paragraph{Case I: FSRQ-Like Powerful Jetted AGNs}
As a benchmark for a quite powerful jet of limited ($\sim 1$kpc) extent, we consider isotropic-equivalent bolometric luminosity $\lbol=10^{48}\ergs$ with an opening angle $\theta_j \sim 18^\circ$
\footnote{This is important to note considering that the true jet power is $\lbol \theta_j^2$}, $\rj = 15$pc ($\hj \sim 1$kpc), and $B_0 = 100\mu$G.  
  Such a large magnetic field is routinely inferred in powerful FR-II jets, but should also pertain to the spines of FR-I jets \citep[e.g.,][]{hardcastle+04,hawley+15}. 
This $\lbol$ prescription is reminiscent of blazars, including FSRQs and BL Lac objects, that have $\lbol$ that go beyond $10^{48} \ergs$ \citep[e.g.][]{ghisellini+09b}.
This should enhance the effects of photodisintegration and the production of neutrinos considering that this $\lbol$ is deemed quite large \citep{ajello+13,tadhunter+16,blandford+19,mingo+19} and particles propagate closer to the base of the jet---where most photons are emitted---compared to the expected size of FR-IIs where $\hj$ can reach hundreds of kpc.

\paragraph{Case II: BL-Lac-like Jetted AGNs}
A jet with a more moderate bolometric luminosity $\lbol = 10^{45}\ergs$ with $\theta_j \sim 18^\circ$, $\rj = 1$pc, and $B_0 = 1.5 $mG is assumed here.
Just as in Case I, the strong magnetic field prescription serves only to study the effect of photodisintegration on UHECRs and the production of astrophysical UHE neutrinos.
We choose to set the jet radius to 1pc to further increase the probability of particle interactions as the bulk of the photon field energy is emitted at the base (See \S\ref{sec:d-dependence} for a discussion on the distance dependence), by setting $\hj$ to the smallest FR-I scales \citep{hawley+15}.

\subsection{UHECR injection spectrum}\label{sec:q}
While the spectrum of UHECRs detected at Earth is measured to be $\propto E^{-2.6}$, the actual spectrum injected by their sources is not well constrained because of the uncertainties in the cosmological distribution of sources and in adiabatic and inelastic losses.
While several authors have considered a spectrum of UHECRs injected into intergalactic space, $E^{-\qi}$, with $\qi = 2$ \citep[e.g.,][]{waxman95,katz+09,aloisio+11}---which may be valid for protons---more recent Auger data favor harder spectra with $1 \leq \qi \leq 1.6$ \citep[e.g.,][]{aloisio+11,gaisser+13,aloisio+14,taylor+15,Auger17c,jiang+21} to explain the observed heavy chemical composition. 

For steeper proton spectra, the rate of injection $\qcr$ has been calculated to be $Ed\qcr/dE(\qi=2) \sim 5 \times 10^{43}$\ergym~for $E \gtrsim 10^{19} $eV \citep[e.g.,][]{katz+09}.
On the other hand, flatter injection spectra would require a slightly larger rate, such that $Ed\qcr/dE(\qi=1) \sim 2 \times 10^{44}$\ergym~for $E > 10^{16} $eV \citep[e.g.][]{aloisio+14}.
The energy generation rate density is moderately affected by the composition, and for different compositions from protons to iron nuclei, the differential energy generation rate is $\approx (0.2-2) \times 10^{44}$\ergym~at $10^{19.5}$~eV \citep{murase+19,jiang+21}.

In this paper, we do not account for propagation effects, and we bracket our ignorance of actual UHECR spectrum by considering injection slopes $1\leq \qi \leq2$.
In the \emph{espresso} framework, the injection spectrum turns out to be flatter than the spectrum of the CR seeds, which should be a power law $\propto E^{-q_{\rm se}}$, with $q_{\rm se}\sim 2-2.7$ \citep{caprioli15}, because reacceleration tends to push particles close to the jet's Hillas limit (\citetalias{mbarek+19}, \citetalias{mbarek+21a}).
It is worth mentioning that $q_{\rm se}$ could approach values of $\sim 2.7$ if we consider relatively short jets like FR-I jets that extend to $\lesssim 10$~kpc, where CRs within the CR halo could be reaccelerated \citep{kimura+18}.
A systematic study of \emph{espresso} acceleration in different kinds of AGN jets is ongoing, but in general we find  that $\qi-q_{\rm se}\sim 0.5-1$, is consistent with the flatter spectra required to explain Auger data \citep[e.g.,][]{aloisio+14,taylor+15}. Similar trends are observed in shear acceleration mechanism, for which it was shown that Auger data can be quantitatively fitted \citep{kimura+18}. 

In general, the spectrum of secondary particles and UHE neutrinos is a function of $q$; therefore we show results for different values of $q_{\rm se}$. In the remainder of this paper, we fix the value of the UHECR injection spectrum q such that $1\leq \qi \leq2$.

\section{Results}\label{sec:uhecrNus}

\subsection{Effects of energy losses on UHECR spectra}
\begin{figure*}
\centering
\includegraphics[width=0.472\textwidth,clip=false,trim= 0 0 0 0]{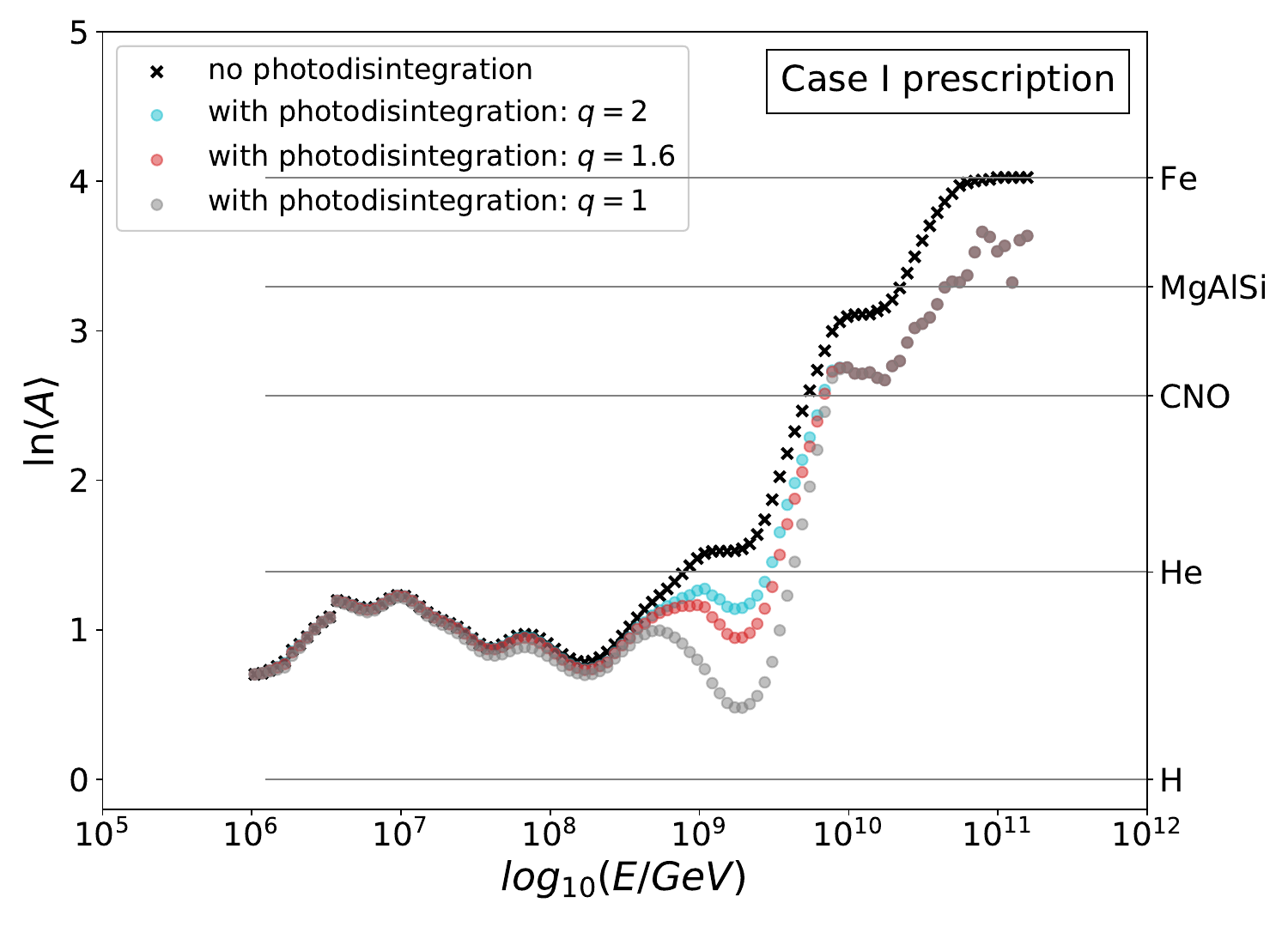}
\includegraphics[width=0.48\textwidth,clip=false,trim= 0 0 0 0]{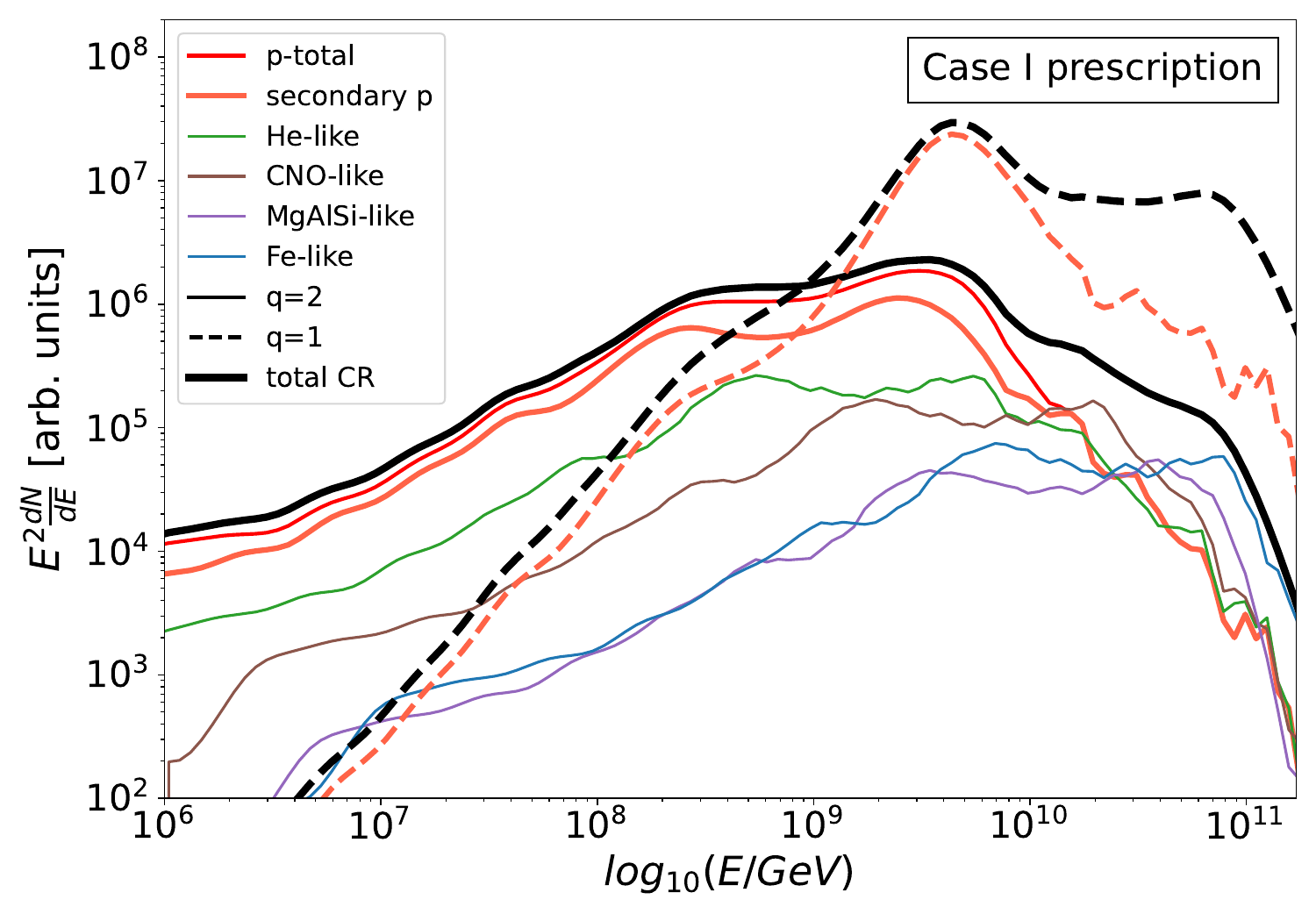}
\caption{Left Panel: Average atomic mass A as a function of energy for the spectrum from the left panel for Case I prescription. The horizontal solid lines correspond to the atomic masses of the injected chemical species. Right Panel: UHECR spectrum including secondary particle spectra. He-like are particles with atomic mass A$ \in [3,8]$; CNO-like with A$ \in [9,18]$; MgAlSi-like with A$ \in [19,35]$; Fe-like with A$ \in [36,56]$. }
\label{A_dependence}
\end{figure*}

Let us start the discussion of our main findings by assessing the role of losses on the spectra of reaccelerated UHECRs.
As discussed above, the parameters for Case I are chosen in order to maximize the potential losses for UHECRs: a powerful, yet compact, source would in fact force particles to propagate closer to the AGN and hence be exposed to the bulk of its non-thermal emission \citep[see, e.g.,][]{dermer07,murase+12}.  
The left panel of Figure~\ref{A_dependence} shows the average atomic mass and spectrum of UHECRs for Case I, with the contribution of different chemical species and for values $\qi=$1, 1.6, and 2. 
It is worth noting that the UHECR spectrum is not significantly affected by photodisintegration, even with a prescription that may magnify its effects because \emph{espresso} acceleration mainly occurs at scales reaching kiloparsec levels, where the photon density becomes lower than that in the blazar region. Figure \ref{tpg} indicates that the optical depth at $\sim10^{20}$~eV is $\sim H_{\rm jet}t_{A\gamma}^{-1}/c\sim10<A$, implying that the effective optical depth---taking into account the inelasticity \citep{murase+10}---is not much larger than unity.  Hence, a significant fraction of the heavy nuclei survives losses because seed reacceleration occurs throughout the jet extent, and not just in the blazar region.
The light component cuts off at a few times $10^{18}$eV, and the overall spectrum gets heavier with increasing energy, consistent with Auger observations \citep{auger14b,auger17,yushkov+19}. 
The right panel of Figure~\ref{A_dependence} breaks down the contribution of each chemical specie for $q=2$ (solid lines, color coded) and also shows the total spectrum for $q=1$ (dotted line).

\citet{unger+15} suggested that heavy ions may be photodisintegrated if the UHECR confinement time were increased around sources due to the presence of magnetic irregularities. In this picture the secondary nuclei originating in these regions would account for the light composition observed below $10^{18}$~eV. Here we observe a similar phenomenology, in the sense that particle scattering in the cocoon produces a light-element bump of reaccelerated secondary protons around $10^{18}$~eV, provided that the seed spectrum is sufficiently flat. This might correspond to the so-called EeV component that is often invoked to fit the low-energy section of the UHECR spectra \citep[e.g,][]{gaisser+13,aloisio+14}.
Note that the position of this bump does not depend on the assumed SGS, in that it corresponds to the Hillas limit for protons; also, increased scattering does not significantly increase the confinement time of particles because particles are more likely to escape sideways from the jet as SGS increases (see \citetalias{mbarek+21a}).

\subsection{Neutrinos from interactions of UHECRs inside their sources}
\subsubsection{UHE neutrino flux expected from a given jetted AGN}\label{uhe-neutrino}

\begin{figure*}
\centering
\includegraphics[width=0.48\textwidth,clip=false, trim= 0 0 0 0]{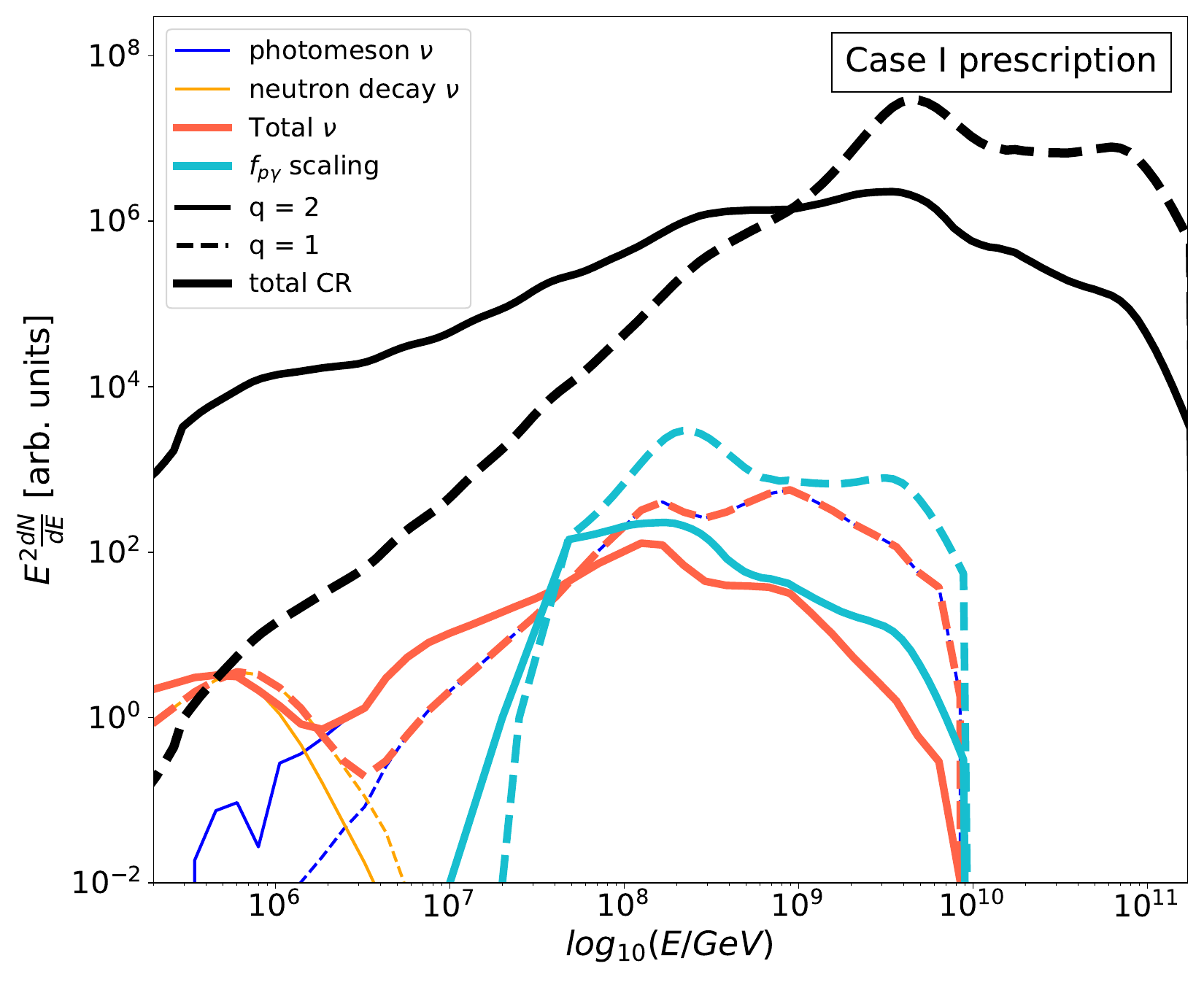}
\includegraphics[width=0.48\textwidth,clip=false, trim= 0 0 0 0]{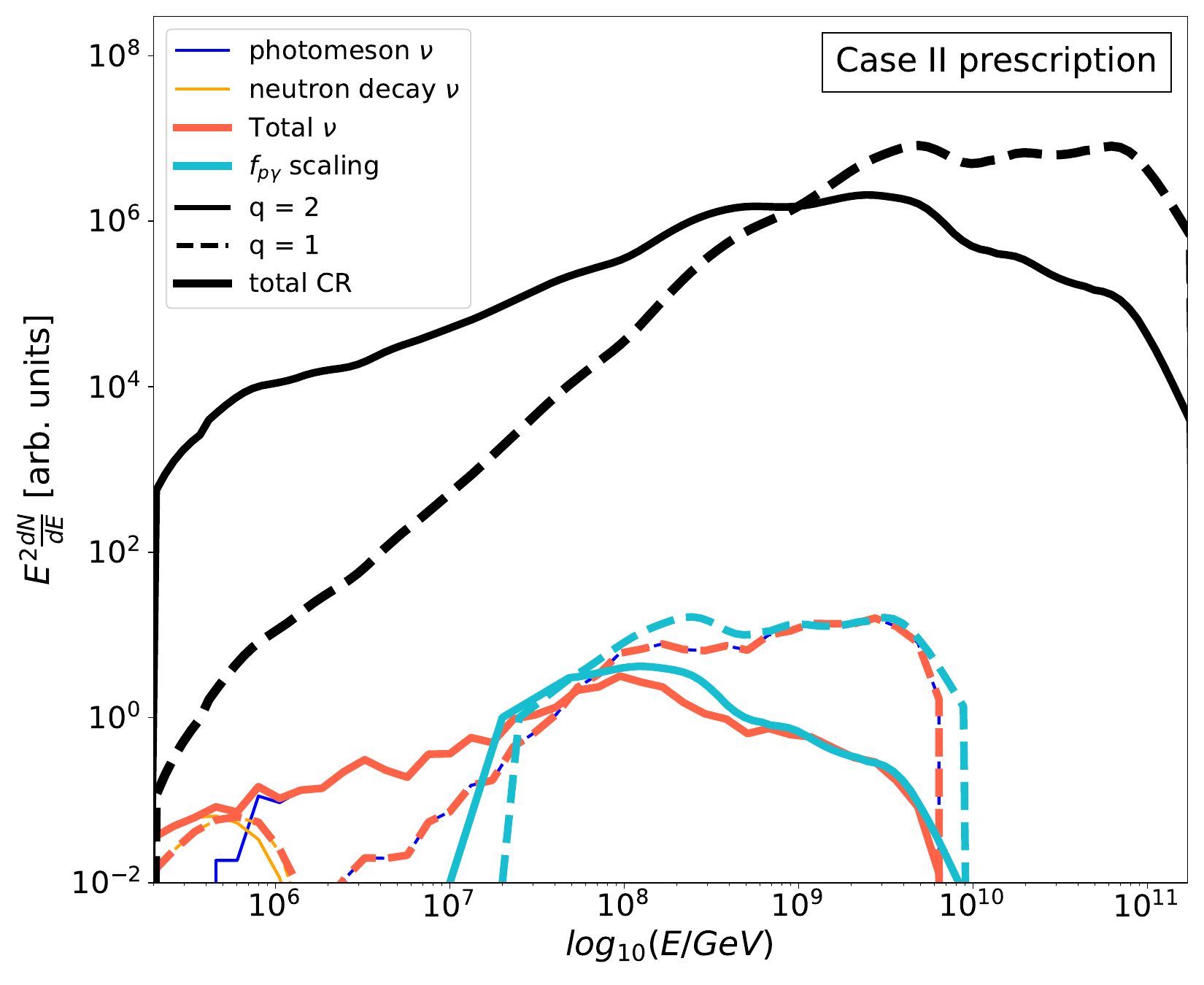}
\caption{Left Panel: Overall CR spectrum including secondary particles in black for the examples of $\qi$= 1, 2 assuming the same energy injection rate at $10^{18}$eV. 
Note that the CR spectrum is not affected by the bolometric luminosity prescriptions because photodisintegration does not play a major role in this energy range. 
Blue and Orange lines: neutrino spectra that ensued from neutron decay, and $p\gamma$ and $pp$ interactions for different radius and bolometric luminosity prescriptions. The spectra are computed based on the methods presented in \S\ref{z_appendix} and \S\ref{sp_appendix}. The teal lines show the expected neutrino based on the $\nu$ scaling from Equation~\ref{eq:scaling}. 
Right Panel: Same as the left panel but for the Case II prescription.}
\label{uhe_nu}
\end{figure*}

The black and red curves in Figure~\ref{uhe_nu} show the expected UHECR and neutrino spectra for two different AGN jets (Case I and II, left and right panel, respectively) and for $\qi=1,2$ (dashed, solid lines), as examples of spectral slopes. Extrapolations to other $\qi$'s are straightforward based on the results in Appendix~\ref{sp_appendix}.

%Three main trends arise, as expected: 
%1) the neutrino flux increases with increasing jet luminosity (compare left and right panels in Figure~\ref{uhe_nu});
%2) flatter injection slopes produce more neutrinos (see Appendix~\ref{sp_appendix});
%3) the highest-energy neutrinos are produced via photomeson production interactions (thin blue curves), while at lower energies ($\lesssim$10~PeV), neutrinos from the neutron decay of photodisintegrated UHECRs dominate the flux (thin yellow curves).
Two main trends arise, as expected: 
1) the neutrino flux increases with increasing jet luminosity (compare left and right panels in Figure~\ref{uhe_nu});
2) the highest-energy neutrinos are produced via photomeson production interactions (thin blue curves), while at lower energies ($\lesssim$10~PeV), neutrinos from the neutron decay of photodisintegrated UHECRs dominate the flux (thin yellow curves).
Finally, we point out that all the neutrinos produced here are from $p\gamma$ interactions; \emph{pp} collisions are negligible and not visible in the plots.
These results do not depend on the level of assumed SGS, since most of the HE neutrinos are produced by powerful AGN jets that can be relativistic even at kiloparsec scales, for which CRs are mainly accelerated via the \emph{espresso} mechanism (\citetalias{mbarek+21a}). We also note that the differences in the total CR spectra between the two panels is quite small for q=2 because the contribution of secondary protons is not as pronounced. 

The red lines in Figure~\ref{uhe_nu} are calculated based on the propagation of test-particles in our fiducial jet, but it is instructive to also calculate the neutrino spectrum resulting from a much simpler approach that ignores the actual jet structure.
Such an analytical \emph{one-zone model} is fully described in Appendix~\ref{nus_appendix}; in a nutshell, a neutrino flux can be estimated via a rescaling of the UHECR flux based on an effective optical depth $f_{p\gamma}$ for $p\gamma$ interactions, with $f_{p\gamma}$ given by:
\begin{equation}\label{eq:scaling}
f_{p\gamma}(\gamma) \sim \sigma_{\rm eff} \xi \frac{L_{\epsilon_{\rm typ}}(\gamma)}{4\pi H_{\rm jet} c}
\end{equation}
where $H_{\rm jet}$ is the extent of the jet, $\epsilon_{\rm typ}\approx 0.5\bar{\epsilon}_{\rm res}/\gamma$ is the most probable photon energy, $\gamma$ is the particle Lorentz factor, $L_{\epsilon}$ is the differential luminosity of the jet at $\epsilon$, $\xi$ is the average energy fraction lost to the pion, and $E_\nu = \alpha E$ (see Appendix~\ref{nus_appendix} and Table \ref{tab:interactions} for more details). 
We can then introduce $f_\nu$ (corresponding to red curves in Figure~\ref{uhe_nu}), which can be thought of as an effective optical depth for UHECR interactions. $f_\nu$ globally captures the order of magnitude of the full kinetic approach and can be used to quickly estimate the expected neutrino flux from a given AGN that is active as an UHECR source.

\subsubsection{Expected flux of UHE source neutrinos}\label{sec:data}
We move now to estimating the overall flux of UHE neutrinos produced by a realistic distribution of the AGN jet luminosity. 
The energy flux in neutrinos that comes from the convolution over the cosmological distribution of their sources can be expressed as \citep[e.g.,][]{murase+14,ahlers+14}: 
\begin{equation}\label{eq:Numain}
\begin{split}
E^2_\nu \Phi_\nu = \frac{c}{4 \pi} \int_0^{z_{\rm max}} \frac{dz}{H(z)} \int d(\ln L) L\frac{d\rho}{dL}(z)\\ \times E_\nu^2 \mathcal{J}_\nu[(1+z)E_\nu,L]
\end{split}
\end{equation}
where $\rho(z)$ is the number density of sources, $L$ is the AGN jets' luminosity, $H(z) = H_0 \sqrt{(1+z)^3 \Omega_M + \Omega_\Lambda}$ is the Hubble parameter, with $\Omega_M \approx 0.3$ and $\Omega_\Lambda \approx 0.7$ for standard $\Lambda$CDM cosmology, and $H_0 = 70$km~s$^{-1}$~Mpc$^{-1}$ is the Hubble constant. 
The $E_\nu^2\mathcal{J}_\nu[(1+z)E_\nu, L]$ term is the average neutrino source luminosity.

In this paper, we consider a scenario in which all the UHECRs are produced in environments with intense photon fields, which should get us closer to an upper limit on the possible neutrino flux. Such a flux must be anchored to some expected UHECR luminosity, hence we scale the $\rho(z) E_\nu^2 \mathcal{J}_\nu$ term with $Ed\qcr/dE(E = 10^{19}\text{eV})$ (see \S\ref{sec:q}), since most CR interactions leading to neutrinos occur at $10^{19}$~eV.
See \cite{murase+10} for upper limits on the diffuse neutrino flux from the sources of UHECR heavy nuclei. 

Both the photon and the UHECR luminosities for an AGN jet should scale with its  bulk power (also see \citetalias{mbarek+19}). 
\citet{ghisellini+09} found that there may be hints that could relate the jet bulk power to accretion disk luminosity; this study was later complemented by findings from \citet{ghisellini+14} where it was asserted that there is a clear correlation between the accretion luminosity and $\gamma$-ray jet luminosity.
The UHECR injection rate at any redshift can then be written as a function of the local one such that: 
\begin{equation}
    E\frac{d\qcr}{dE}(z, q) = E\frac{d\qcr}{dE}(z= 0, q) \zeta_z(\lx, z),
\end{equation}
where $\zeta_z$ is a cosmological evolution factor defined in equation~\ref{eq:fz} of Appendix~\ref{cosmo_appendix} and $\lx$ the X-ray luminosity of the AGN sources.  
The prescribed photon fields in our framework are related to $\lx$ such that $\lbol / \lx  = 10^{4.21}$ where the constant of proportionality is obtained by modeling the $\gamma$-ray luminosity function through the observed X-ray luminosity \citep[see, e.g.,][]{ueda+03,inoue+09b,harding+12}. This is broadly consistent with the more recent results on the blazar luminosity function \citep{Ajello12,Ajello14}.

Assuming that there are different classes of jetted AGNs, we can then rewrite Equation~\ref{eq:Numain} as:
\begin{equation}\label{eq:Numain2}
\begin{split}
    E^2_\nu  \Phi_\nu=\frac{c}{4 \pi} \int_0^{z_{\rm max}} \frac{dz}{H(z)} E\frac{d\qcr}{dE}(z= 0,\qi) \\ \times \sum_i w_{i} f_\nu(L_{i}, E, \qi) \zeta_z(10^{-4.21}L_{i}, z) 
\end{split}
\end{equation}
where the weights $w_i$ of the different classes are defined as the relative X-ray injection in Mpc$^{-3}$ of each AGN type:
\begin{equation}\label{eq:weight}
    w_{\rm log(\lbol)} = w(\lx) = \frac{ \lx \frac{d\psi}{d\lx }   }{ \sum_{k} L_{ k} \frac{d\psi_{k}}{dL_{ k}}    }
\end{equation}
such that $\frac{d\psi}{d\lx}(\lx, z)$ is the $z$-dependent X-ray luminosity function of the AGN sources per luminosity per comoving volume (see \citetalias{murase+14} for more details).
As discussed in \S\ref{sec:q}, $Ed\qcr/dE$ depends on the injection slope $\qi$, and in particular we have that $Ed\qcr/dE(z=0,\qi=2) \sim 5 \times 10^{43} \ergym$ \citep{katz+09} at $\sim 10^{19}$~eV generally assumed for pure proton compositions. 
For heavier UHECR compositions, a slightly larger injection could be favored such that $Ed\qcr/dE(z=0, 0.1\leq \qi \leq 2.7) \sim (0.3-2) \times 10^{44} \ergym$ at $\sim 10^{19}$~eV \citep{jiang+21}. 
Integrated values of the energy generation rate density have also been reported such that $\qcr(z=0,\qi \leq 1.6) \sim 2 \times 10^{44} \ergym $ for $E> 10^{16}$~eV/n \citep{aloisio+14} and $\qcr(z=0,\qi =1) \sim 5 \times 10^{44} \ergym $ at $\sim 10^{19}$~eV \citep{Auger17c}. Here, we assume a rate equivalent to $Ed\qcr/dE(z=0,\qi < 2) = 2 \times 10^{44} \ergym$ and $Ed\qcr/dE(z=0,\qi=2) = 0.5 \times 10^{44} \ergym$ at $10^{19}$~eV to maximize UHECR injection.

In this study, we consider contributions from two types of AGN jets with two different isotropic-equivalent bolometric luminosities: 
i) FSRQ-like powerful jetted AGNs with $\lbol=L_{48} \sim 10^{48}\ergs$ (Case I), and ii) BL-Lac-like jetted AGNs with $\lbol = L_{45} \sim 10^{45}\ergs$ (Case II). 
We note that the associated $\lx \frac{d\psi}{d\lx } \approxprop \lx^{-2}$, so the relative contribution of each AGN type in this case is roughly the same, i.e., $w_{45} \sim w_{48}$ and the total energy injected per unit time in UHECRs is comparable in our case (see discussion below). 
Note that they could, in principle, accelerate particles to different maximum energies (see the discussion in \citetalias{mbarek+21a}).
We finally obtain:
\begin{equation}\label{eq:Numain3}
\begin{split}
    E^2_\nu  \phi_\nu \approx \frac{c}{4 \pi} \int_0^{z_{\rm max}} \frac{dz}{H(z)} E\frac{d\qcr}{dE} (\qi) \\ \times \Big[ w_{45} f_\nu(L_{45}, E, \qi) \zeta_z(10^{-4.21}L_{45}, z)  \\ + w_{48} f_\nu(L_{48}, E, \qi) \zeta_z(10^{-4.21}L_{48}, z) \Big]
\end{split}
\end{equation}

\begin{figure}
\centering
\includegraphics[width=0.48\textwidth,clip=false, trim= 0 0 0 0]{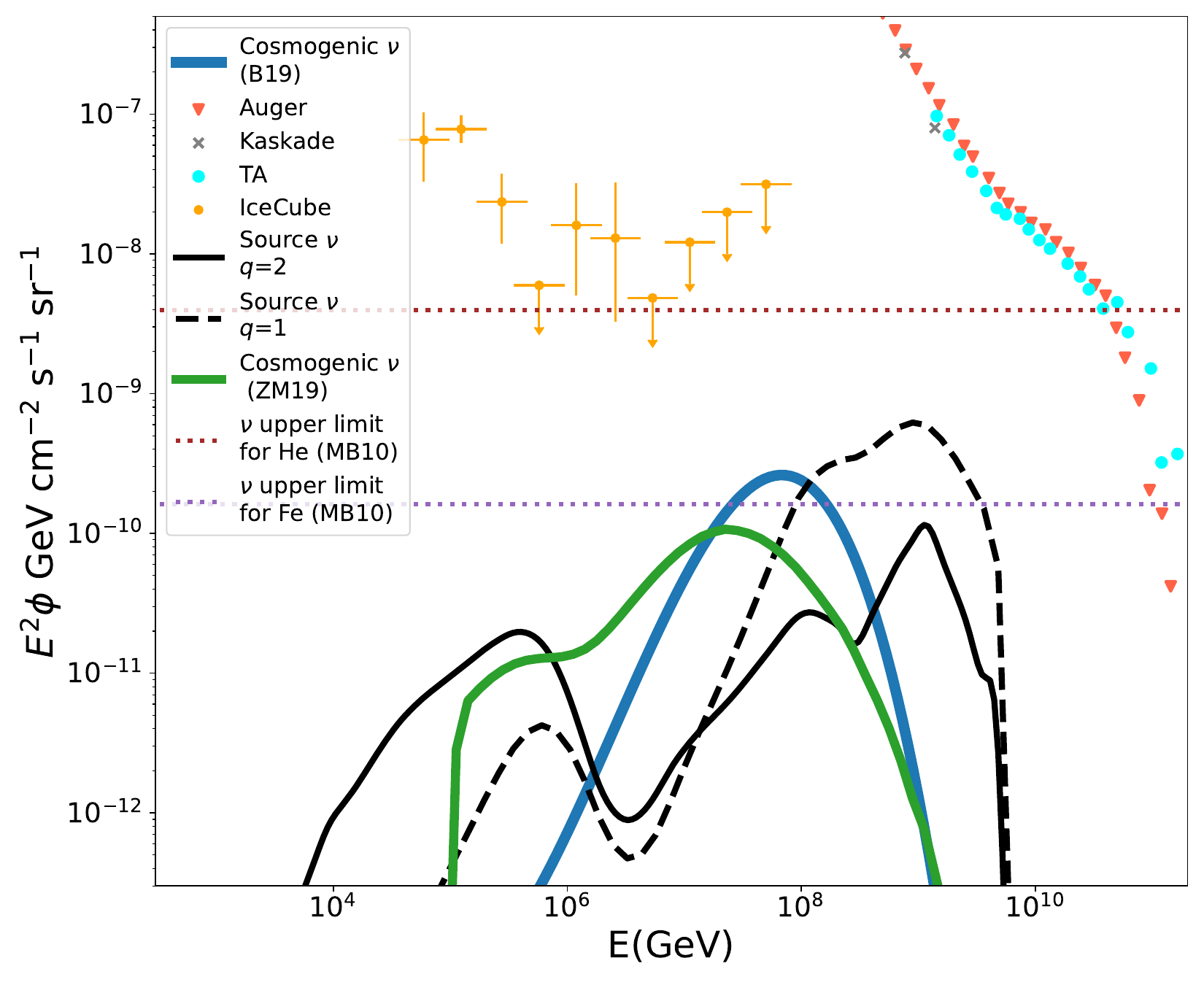} 
\caption{Expected upper bounds for the source neutrino flux (black) from UHECR interactions for three different injection slopes q=2 (solid) and q = 1 (dashed) including cosmological effects. This flux is compared to the expected cosmogenic neutrino fluxes  \citep[green line from][]{zm19} for an ion composition similar to ours  and a more proton rich composition \citep[blue bands from ][]{batista+19} modeled according to an AGN source evolution. The dotted lines show nucleus-survival bounds on the diffuse neutrino flux from sources of UHECR heavy nuclei for Helium and Iron based on analytical calculations in \citet{murase+10} with $Ed\qcr/dE(z=0,\qi=2)=0.6 \times 10^{43} \ergym$ and $\xi_z=7.2$.
IceCube neutrino data, along with UHECR data from Auger, KASCADE, and TA are also included for reference.}
\label{uhe_nu-data}
\end{figure}

%\paragraph{Resulting neutrino flux}
Figure~\ref{uhe_nu-data} shows the resulting UHE source neutrino fluxes (black lines) based on Equation~\ref{eq:Numain3} and expected cosmogenic neutrino fluxes (green and blue lines). The green line is the expected cosmogenic neutrino flux from radio galaxies assuming an ion composition consistent with ours \citep[taken from][where the shear acceleration scenario following \cite{kimura+18} is considered]{zm19}. The blue bands, on the other hand, show the expected cosmogenic flux for an AGN model that is more proton-rich based on models that fit Auger's spectral features with different confidence levels \citep{batista+19}. The pink and brown dotted lines show nucleus-survival bounds on the diffuse neutrino flux from sources of UHECR heavy nuclei \citep{murase+10}, which is consisten with the intuition that neutrino production is inefficient in photon-poor environments allowing the survival of nuclei.
The yellow data points show the fluxes of the high-energy neutrino events \citep{kopper17}.
Shown are also the CR data from Auger \citep{AUGER20}, Telescope Array \citep{TA16}, and KASCADE-GRANDE \citep{KASCADE13b} for reference.

A few points are worth noticing.
\begin{itemize}
    %\item At a fixed UHECR injection rate, the normalization of the source neutrino flux is strongly dependent on the spectral slope of their parent UHECRs such that $f_\nu(E_\nu)/f(E_\nu)= \alpha^{q-1}$ with $E_\nu = \alpha E$ (See Appendix~\ref{sp_appendix} and Table~\ref{tab:interactions} for more details).
    %The flatter the injection spectra, the more neutrinos are produced because the more power is available in CRs with energies above $10^{18}$~eV.
    \item If injection spectra are sufficiently flat, the flux of source neutrinos may dominate the expected flux of cosmogenic neutrinos \citep[see also][]{murase+14,rodrigues+21}\footnote{Although beamed neutrino emission is assumed in the previous studies, which is different from our results where the neutrino emission is nearly isotropic (see \S~\ref{sec:ang} for more details)}.
    Even for a moderately flat spectrum of $q=2$, their relative contribution at $\sim 10^{17}$~eV turns out to be comparable.
    \item Given the dependence of the source neutrino flux on $q$ and $\alpha$, neutrinos from the $\beta$ decay of photodisintegration byproducts might become a non-negligible fraction of the flux observed by IceCube below a few PeV, although their fluxes are likely to be lower than those from the photomeson production \citep[see also][]{murase+10}.
    \item The neutrino fluxes that we obtain for q=2 are lower than the nucleus-survival bounds on the diffuse neutrino flux stemming from UHECRs with $A \geq 4$.
\end{itemize}

\subsubsection{Upper limits of the UHE neutrino flux in the reacceleration sceanrio}\label{sec:upperLimit}
The results shown in Figure~\ref{uhe_nu-data} should be regarded as close to \emph{upper limits} of the neutrino flux from AGN jets since we assumed that all UHECRs are produced in jetted AGNs with considerable isotropic-equivalent bolometric luminosities (between $10^{45}$ and $10^{48} \ergs$).

We remark that:
\begin{itemize}
    \item The jet simulations that we consider have a limited extent---consistent with an FR-I jet---with a bolometric luminosity that reaches that of the most powerful blazars. This further maximizes the effects of the photon fields that are generated close to the base of the jet.
    \item Relaxing any of our assumptions would reduce the  contribution of the non-thermal photon background to $p\gamma$ collisions; the CMB, BLR, and dusty torus contributions alone would provide a source neutrino flux that would be a few orders of magnitude smaller (see left panel of Figure \ref{tpg}), subdominant with respect to both the cosmogenic neutrino flux and the astrophysical neutrino flux in the IceCube band.
    \item  Our results are sensitive to the luminosity function as that directly affects UHECR normalization, the effect of photodisintegration, and the neutrino flux prediction. For example, lower-luminosity AGNs (and/or other classes of sources) 
    may give more contributions to the observed UHECR flux, e.g., via the shear acceleration mechanism, because their luminosity density is larger \citep{Ajello14}.
    Then the expected neutrino flux would be reduced because the expected neutrino power scales with the bolometric luminosity as shown in Figure~\ref{uhe_nu}. Our neutrino flux should still be regarded as an upper limit, since including lower-energy AGNs should not result in a higher neutrino flux.
\end{itemize}

The results of Figure~\ref{uhe_nu} and~\ref{uhe_nu-data} lead to two very general considerations. 
On one hand, neutrinos produced by the $\beta$ decay of secondary neutrons should be accounted for when calculating the flux of neutrinos in the PeV range. 
%\km{We found that our numerical fluxes are consistent with the analytical estimates. For example using Eq.~(15) with the , we expect that }
On the other hand, in order to explain the whole IceCube flux with neutrinos from UHECR sources, photodisintegration would need to happen at a much higher rate than what is estimated here, which would be inconsistent with the presence of heavy elements at the highest energies. 
In other words, the optical depth $f_{\nu}$ required to produce a sizable source neutrino flux would necessarily lead to the complete photodisintegration of heavy UHECR nuclei; this result is general and independent of the UHECR source or acceleration mechanism.
Overall, better UHECR chemical composition measurements and constraints on the UHECR injection slopes may pose more stringent limits on the expected contribution of neutron-decay neutrinos to the observed IceCube flux.

\subsubsection{Sites of production and angular distribution of escaping neutrinos}\label{sec:ang}
\begin{figure}
\centering
\includegraphics[width=0.48\textwidth,clip=false, trim= 0 0 0 0]{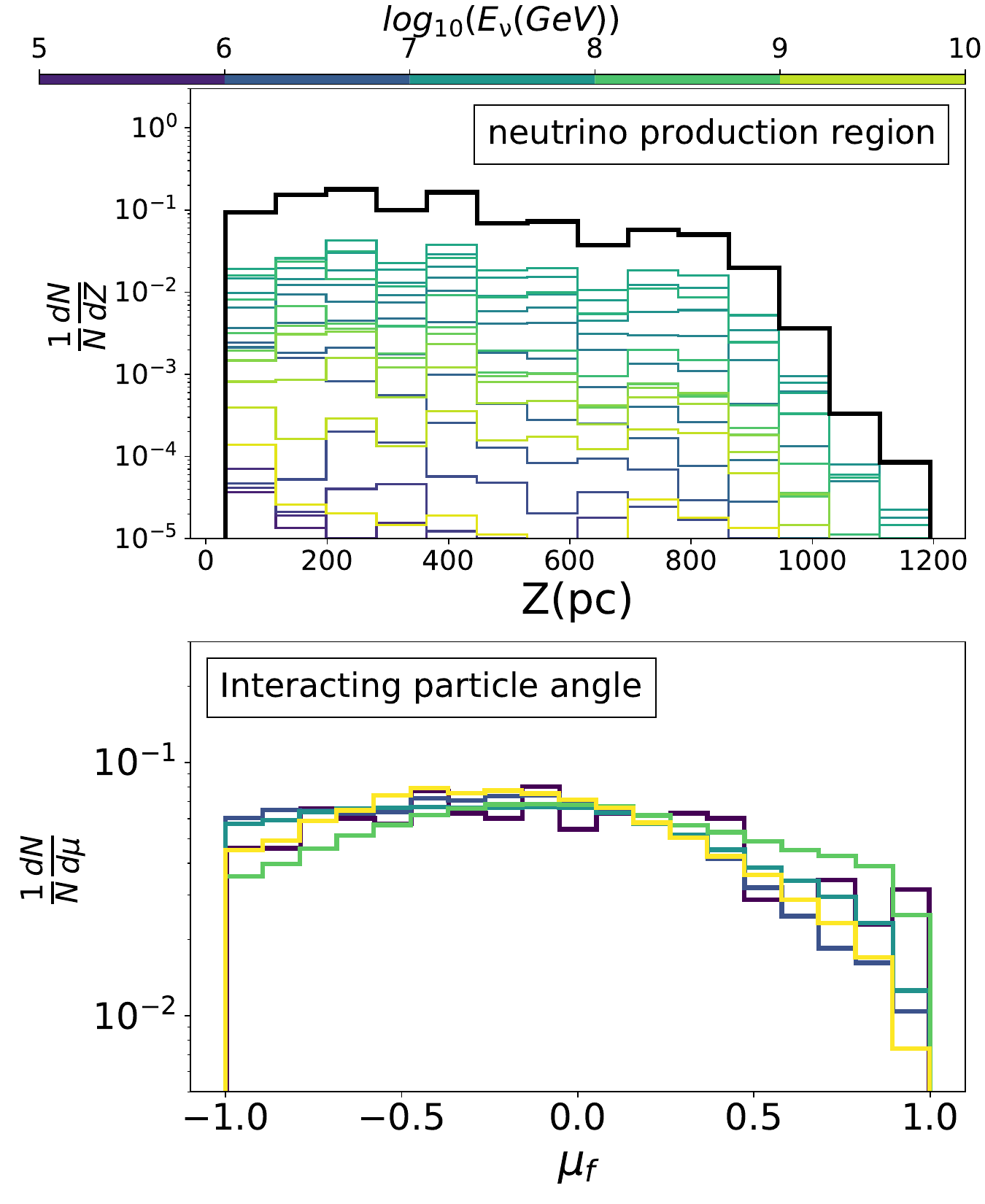}
\caption{Upper Panel: Final Z position of neutrinos for the Case I prescription to trace the regions where neutrinos are most likely to be produced. Neutrinos are preferentially produced close to the base of jet within $\sim 800$pc, but a non-negligible fraction of neutrinos is produced at larger distances.
Lower Panel: Distribution of the final angles of escaping neutrinos where $\mu_f $ is the cosine of the angle between the particle momentum and the jet axis. We can see that neutrinos escape the jet ``quasi-isotropically''.}
\label{neutrino_geometry}
\end{figure}

In our simulations we keep track of where neutrinos are produced and in which direction they escape.
The top panel of Figure \ref{neutrino_geometry} shows that a considerable fraction of neutrinos is produced at relatively large distances ($\sim 1$ kpc) away from the base of the jet as a result of two competing effects:
on one hand, the photon field intensity declines as $D^{-2}$, and 
on the other hand, the non-thermal emission cone affects a greater volume as we move away from the base of the jet. 
We observe similar trends for all of our prescriptions, so we argue that the photodisintegration of heavy UHECR nuclei and neutrino production mainly occur at intermediate distances, not too close to the blazar region but well before the jet's end.

The bottom panel of Figure \ref{neutrino_geometry}, instead, shows the distribution of the cosines of the final angles of flight of the neutrinos produced through $p \gamma$ interactions and neutron decay for Case I.
Neutrinos are released quasi-isotropically, essentially because UHECRs are efficiently isotropized in the cocoon (see also \citetalias{mbarek+19} and \citetalias{mbarek+21a}).

This is important for neutrino astronomy because any correlation between neutrino directions of arrival and the AGN population is intrinsically connected to how neutrinos are released from their sources \citep[see also the discussion in][]{caprioli18}. 
If neutrinos are strongly beamed along the jet axis, then they would preferentially be associated with AGN jets that point towards us such as BL Lac objects and FSRQs (i.e., blazars). 
On the other hand, if the emission were less beamed, we might expect all radio-loud AGNs to contribute to the flux, generally producing a more isotropic signal. 
Our results suggest that we should receive neutrinos from non-blazar AGNs, too, which makes it harder to assess AGNs as UHE neutrino sources \emph{on a statistical basis}. This prediction is different from the neutrino emission model postulated by \citet{rodrigues+21}. This, however, does not preclude the association of individual events with given sources.

\section{Conclusions}\label{sec:conc}
In this paper, we studied the effects of photodisintegration on UHECRs accelerated in powerful AGN jets and estimate the ensuing flux of high-energy neutrinos. 
Particles are propagated in a mechanism-agnostic way and we find that UHECRs are generically accelerated via the \emph{espresso} reacceleration of galactic-like CRs \citep{caprioli15}---independently of the jet morphology (\citetalias{mbarek+19}) and on the details of particle transport (\citetalias{mbarek+21a})---in promising ``relativistic'' jets, especially for powerful jetted AGN such as FR-II galaxies.

We used a bottom-up approach in which test particles (CR seeds) are propagated in a fiducial 3D MHD simulation of an ultra-relativistic jet, augmented with realistic modeling (à la \citetalias{murase+14}) of the photon fields responsible for UHECR losses and hence for neutrino production via different channels (see Table \ref{tab:interactions}).
We considered different prescriptions for the AGN size and luminosity (\S\ref{jet-lum}) to maximize the interaction rate and, hence, give an optimistic estimate on the expected source neutrino flux, under the additional assumption that powerful jetted AGNs are responsible for the total flux of UHECRs on Earth. We note that relaxing any of these hypotheses would generally lead to a lower neutrino flux.

To summarize, the main findings of this paper are as follows:
\begin{enumerate}
    \item For typical baryon densities and photon fields, $pp$ interactions are negligible with respect to $p\gamma$ collisions; moreover, for powerful jetted AGN the most relevant photon field is provided by the non-thermal jet emission, which dominates over the IR dusty torus emission, the optical stellar light, and even the CMB.
    \item UHECRs are not heavily affected by photodisintegration, even in the most luminous AGNs; 
    the spectrum of the highest-energy particles gets heavier with increasing energy, as reported by Auger \citep{auger14a}. This situation is different from that in the more compact blazar region \citep{murase+12,murase+14,rodrigues+21}. 
    \item In general, the expected neutrino flux scales with the AGN luminosity and with the slope of the seed and the reaccelerated CR spectra (Figure \ref{uhe_nu});
    the rather hard spectra ($q\lesssim 1.5$) required to explain Auger data \citep[see, e.g.,][]{aloisio+14, taylor14} would maximize the neutrino yield $\geq 10^{17}$eV with respect to softer UHECR spectra.
    \item For the most optimistic scenarios (flat UHECR injection spectra), UHE neutrinos produced inside AGN jets may dominate, or at least be comparable to, the expected flux of cosmogenic neutrinos produced during UHECR propagation across the universe (Figure \ref{uhe_nu-data}).
    \item Even if AGNs sustained the totality of the UHECR luminosity of the universe, their steady neutrino emission could not account for the entire IceCube astrophysical neutrino flux (Figure \ref{uhe_nu-data}).
    However, a non-negligible fraction of the source neutrinos in the IceCube band may come from the $\beta$-decay of photodisintegration byproducts, rather than from photomeson production.
    \item In the reacceleration scenario, neutrinos should be released quasi-isotropically which strongly suggests that non-blazar jets that correspond to radio galaxies should contribute to the UHE astrophysical neutrino flux. This prediction is different from models of beamed neutrino emission \citep[e.g.,][]{rodrigues+21}. 
\end{enumerate}

Note that CR acceleration and neutrino production may occur in different regions. 
Large-scale jets eventually become subrelativistic, where UHECRs may be accelerated by the stochastic acceleration mechanism \citep{kimura+18,rieger19}. 
However, the photon density is so low that neutrino production there was shown to be inefficient \citep{zm19}. On the other hand, AGNs with large-scale jets are embedded in galaxy groups and clusters, where the cosmic rays escaping from the jets can be confined in the intracluster medium for a cosmological time, in which high-energy neutrinos are produced predominantly via $pp$ interactions \citep{fang+18}. We defer to a future study the estimate of the neutrino flux expected from nearby sources, such as M87 and Centaurus A, which may be detectable by current/future experiments even if such moderately powerful AGNs were not contributing substantially to the observed flux of UHECRs.

\begin{acknowledgements}
We thank the anonymous referee for providing insightful comments that helped us to correct the neutrino flux from neutron decay in Figure \ref{uhe_nu-data}. Simulations were performed on computational resources provided by the University of Chicago Research Computing Center.
We thank B. Theodore Zhang for providing us with tables of the cosmogenic neutrino flux. This work of D.C. was partially supported by NSF through grant No.~PHY-2010240, while the work of K.M. was supported by the NSF Grants No.~AST-1908689, No.~AST-2108466 and No.~AST-2108467, and KAKENHI No.~20H01901 and No.~20H05852.
\end{acknowledgements}

\appendix
\section{Spectra of nuclei}\label{z_appendix}
We propagate particles of given rigidity $R$, which for different species with different atomic charge $Z$ corresponds to an energy $E = RZ$.
The normalizations of the different ion species are chosen according to the abundance ratios at $10^{12}$ eV such that $K= K_{\rm se}/K_H \sim$ [1, 0.46, 0.30, 0.14] for He, CNO, MgAlSi and Fe respectively. Hence:
\begin{equation}\label{rig1}
f(E_i) = K f(ZR)
\end{equation}
With $f(R) = f_0 R^{-q}$ where $\qi$ is the spectral slope. Then: 
\begin{equation}\label{rig2}
f(E_i) = f_0 K Z^{-\qi } \big( E_i \big)^{-\qi}
\end{equation}
where $E_{\rm i}$ is the ion energy and R is the rigidity. 

\section{Photomeson production interactions}\label{photomes_appendix}
We introduce $t_{p \gamma}$ as the cooling time of a proton with energy $E_p$ due to photomeson interactions, i.e.:
\begin{equation}
    t_{p \gamma}(E_p) = \frac{E_p}{dE_p/dt} \sim \frac{E_p}{\Delta E_p/\Delta t}.
\end{equation}

We follow \citet{stecker68} and for an isotropic photon field we have:
\begin{equation}\label{s0}
    t^{-1}_{p \gamma}(\gamma_p) \approx \frac{\xi c}{2 \gamma_p^2} \int^{\infty}_{\bar{\epsilon}_{\rm th}/2\gamma_p} d\epsilon n_\gamma(\epsilon) \epsilon^{-2}  \int^{2 \gamma_p \epsilon}_{\bar{\epsilon}_{\rm th}}  \epsilon ' \sigma(\epsilon ') d\epsilon ',
\end{equation}
where $\epsilon$ is the photon energy in the black hole frame, $\epsilon '$ its energy in the proton frame, $\xi$ is the inelasticity, and $\bar{\epsilon}_{\rm th} \approx 0.15$ GeV is the threshold energy in the proton frame.
We integrate over $\epsilon ' = \gamma_p \epsilon(1 - \beta \cos{\theta})$, where $\theta$ is the angle between the particle momentum vectors in the black hole frame and $\beta \approx 1$. 
For $0 \leq \epsilon ' \leq 2 \gamma_p \epsilon$, we can introduce the effective cross section $\sigma_{\rm eff} = 70 ~\mu$b \citep{dermer+14} with $\sigma_{p\gamma}(\bar{\epsilon})\sim \sigma_{\rm eff}H(\bar{\epsilon}-\bar{\epsilon}_{\rm th})$ and obtain:

\begin{equation}\label{s3}
    t^{-1}_{p \gamma}(\gamma_p) \sim \sigma_{\rm eff} \xi c \int^{\infty}_{\bar{\epsilon}_{\rm th}/2\gamma_p} d\epsilon n_\gamma(\epsilon, d_p) [1 - \frac{\bar{\epsilon}_{\rm th}^2}{4 \epsilon^2 \gamma_p^2}],
\end{equation}
where $d_p$ expresses the spatial dependence of $n_\gamma$. 

When considering interactions between the photons and a nucleus of atomic number $A$ and energy $E_A$, we assume $\sigma_{\rm eff,A} = A\sigma_{\rm eff}$ and $\xi_A = \xi/A$. 
Based on equation~\ref{s3}, the cooling time for nuclei $t^{-1}_{N \gamma}$ eventually reads:
\begin{equation}\label{s3n}
    t^{-1}_{N \gamma}(\gamma_A) \sim \sigma_{\rm eff} \xi c \int^{\infty}_{\bar{\epsilon}_{\rm th}/2\gamma_A} d\epsilon n_\gamma(\epsilon, d_p) [1 - \frac{\bar{\epsilon}_{\rm th}^2}{4 \epsilon^2 \gamma_A^2}]; \quad \gamma_A \equiv \frac{E_A}{Am_pc^2}.
\end{equation}

Non-thermal emission in jets is typically beamed;
hence, the angle $\theta$ between the proton momentum and the target photon is fixed and we can express $t^{-1}_{p \gamma}(\gamma_p)$ for an individual proton interacting with a beamed photon field as:
\begin{equation}\label{pg-angle}
     t^{-1}_{p \gamma}(\gamma_p) = \xi c (1- \beta \cos{\theta}) \int^{\infty}_\frac{{\bar{\epsilon}_{\rm th}}}{\gamma_p(1- \beta \cos{\theta})} d\epsilon \sigma_{p \gamma}(\epsilon')  n_\gamma(\epsilon) =
     \sigma_{\rm eff}\xi c(1- \beta \cos{\theta})   \int^{\infty}_\frac{{\bar{\epsilon}_{\rm th}}}{\gamma_p(1- \beta \cos{\theta})} d\epsilon  n_\gamma(\epsilon)
\end{equation}
such that again $\epsilon' = \epsilon \gamma_p(1- \beta \cos{\theta})$. 

\section{Photodisintegration Interactions due to the Giant Dipole Resonance}\label{photodis_appendix}
We follow \citet{murase+10} in calculating the  photodisintegration interaction time $t_{A\gamma}$ (for an isotropic background), obtaining an expression similar to equation~\ref{s0}, with $\gamma$ the ion Lorentz factor:
\begin{equation}\label{A0}
    t^{-1}_{A \gamma}(\gamma) \approx \frac{ c}{2 \gamma_A^2}  \int^{\infty}_{\bar{\epsilon}_{\rm th}/2\gamma} d\epsilon n_\gamma(\epsilon, d_p) \epsilon^{-2} \int^{2 \gamma \epsilon}_{\bar{\epsilon}_{\rm th}}  \sigma_{\rm A\gamma}(\epsilon ') \epsilon ' d\epsilon ';
\end{equation}
here $\sigma_{\rm A\gamma}$ is the GDR photodisintegration total cross section, which reads \citep{karakula+93}:
\begin{equation}\label{sigag}
    \sigma_{\rm A\gamma}(\varepsilon') = \frac{\sigma_G  \epsilon'^2 (\Delta \epsilon_{\rm G}')^2}{ [\epsilon_{\rm G}'^2 - \epsilon'^2]^2 + \epsilon'^2 (\Delta \epsilon_{\rm G}')^2}
\end{equation}
where $\sigma_G = 1.45 \times 10^{-27}A$cm$^2$, $\epsilon_{\rm G}' = 42.65A^{-0.21}$ MeV $(0.925A^{2.433}$ MeV) for $A>4$ $(A\leq 4)$, and $\Delta \epsilon_{\rm G}' \sim 8$MeV. 

For isotropic photon spectra, we can simplify equation \ref{sigag} by posing $\sigma_{\rm A\gamma} \sim \sigma_{\rm G} \Delta \epsilon_{\rm G}' \delta (\epsilon ' - \epsilon_{\rm G}')$. Equation~\ref{A0} becomes:
\begin{equation}\label{eq:phot}
    t^{-1}_{A \gamma}(\gamma) \sim \frac{c \sigma_{\rm G}}{2 \gamma_A^2} \Delta \epsilon_{\rm G}'  \int^{\infty}_{\bar{\epsilon}_{\rm th}/2\gamma} d\epsilon n_\gamma(\epsilon, d_p) \epsilon^{-2} \int^{2 \gamma \epsilon}_{\bar{\epsilon}_{\rm th}}   \delta (\epsilon ' - \epsilon_{\rm G}') \epsilon ' d\epsilon ' \sim \frac{c \sigma_{\rm G}}{2 \gamma_A^2} \Delta \epsilon_{\rm G}'   \epsilon_{\rm G}' \theta(\epsilon_{\rm G}' - \bar{\epsilon}_{\rm th})  \Xi(d_p)
\end{equation}
where $\gamma_A = E_A/(m_Ac^2)$, with
\begin{equation}
% \[
    \theta(x)= 
\begin{cases}
    0,& \text{if } x < 0\\
    1/2,& \text{if } x = 0\\
    1, & \text{if } x > 0
\end{cases}
%\]   
\end{equation}

We assume that $\epsilon_{\rm G}' - \bar{\epsilon}_{\rm th}>0$ where $\bar{\epsilon}_{\rm th}$ is the threshold energy (10 MeV for photodisintegration) and $\epsilon_{\rm G}'$ is the energy at which the cross section peaks. 
Also, $\Xi(d_p) = \int^{\infty}_{\bar{\epsilon}_{\rm th}/2\gamma} d\epsilon n_\gamma(\epsilon, d_p) \epsilon^{-2} \theta(2 \gamma \epsilon - \epsilon_{\rm G}')$. For example, if we deal with two prominent BLR  emission lines $\epsilon_{\rm HI} = $ 10.2 eV and $\epsilon_{\rm  Ly\alpha} = $40.8 eV, then:

\begin{equation}\label{eq:phot-angle}
    \Xi(d_p) = \frac{n_\gamma(\epsilon_{\rm HI}, d_p)\theta(2 \gamma \epsilon_{\rm HI} - \epsilon_{\rm G}')}{\epsilon_{\rm HI}^2} + \frac{n_\gamma(\epsilon_{\rm Ly\alpha}, d_p)\theta(2 \gamma \epsilon_{\rm Ly\alpha} - \epsilon_{\rm G}')}{\epsilon_{\rm Ly\alpha}^2}
\end{equation}

For a nucleus interacting with a beamed photon field, the angle $\theta$ between the photon and nucleus momenta is known and we can write $t^{-1}_{A \gamma}(\gamma_p)$ as:
\begin{equation}\label{}
     t^{-1}_{A \gamma}(\gamma_A) = c (1- \beta \cos{\theta}) \int^{\infty}_\frac{{\bar{\epsilon}_{\rm th}}}{\gamma_A(1- \beta \cos{\theta})} d\epsilon \sigma_{A \gamma}(\epsilon')  n_\gamma(\epsilon)
\end{equation} 
where $\sigma_{A \gamma}(\epsilon')$ is given by equation \ref{sigag}.

\section{Spectra of secondary nuclei}\label{sp_appendix}
The spectra of secondary nuclei are intrinsically dependent on the slopes of primary spectra, which in turn depend on the slope of the seed spectrum. Let us consider primary spectra as power laws in energy $f(E) = f_0 E^{-q}$. For the photodisintegration process, under the GDR approximation, the particle number conservation is justified, in which we get: 
\begin{equation}
f(E) dE = f'(E') dE'
\end{equation}
where $E$ is the energy of the primary nucleus and $E' = \alpha E$ is the energy of the secondary particle. For example, for a secondary nucleus we have $\alpha\approx(A-1)/A$ while $\alpha\approx1/A$ for a secondary nucleon.

Then $dE' = \alpha dE$ and we obtain:
\begin{equation}
     f'(E') =  f(E) \frac{dE}{dE'} = f_0 \frac{E^{-q}}{\alpha}
\end{equation}
and therefore:
\begin{equation}
    f'(E') = f_0 (E')^{-\qi} \alpha^{\qi- 1} = f_0  E'^{-q} \alpha^{\qi - 1}
\end{equation}
And finally:
\begin{equation}\label{rat1}
    f_s(E_s)/f(E_s)= f_0 E_s^{-q} \alpha^{q-1}/f(E_s) = \alpha^{q-1}
\end{equation}
where $E_s$ is the energy of the secondary nucleus or nucleon.
Eventually, the relative normalization of primary and secondary spectra reads $f_s(E_s)/f(E_s)= \alpha^{q-1}$.

\section{Spectra of secondary neutrinos and scaling}\label{nus_appendix}
We consider the photomeson optical depth $f_{p\gamma}$ --- a proxy for the neutrino production efficiency produced through photomeson production interactions---to estimate the expected astrophysical UHE neutrino spectrum based on the average bolometric luminosity and extent of the jet in the black hole frame. 
Primary spectra are expressed as power laws in energy $f(E) = f_0 E^{-q}$, and as a result, the number of primary particles is not the same as that of secondary particles. Using $f_{p\gamma}$ we write
\begin{equation}
E_\nu^2 f_\nu(E_\nu) \approx \frac{3}{8} f_{p\gamma}(E) E^2 f(E),  
\end{equation}
where $E$ is the energy of the primary particle, $E_\nu= \alpha E$ is the typical energy of the secondary neutrino, and $\alpha$ can be read in the last column of Table~\ref{tab:interactions} for the different processes that yield neutrinos. While computing the expected neutrino  flux, we need to account for a factor of $1/2$ that comes from the charged to neutral pion ratio, and another factor of $3/4$ stemming from pion and muon decay.  
Then we have:
\begin{equation}\label{rat2}
    f_\nu(E')/f(E')\approx \frac{3}{8} f_{p\gamma}(E'/\alpha) \alpha^{q-2}
\end{equation}

Since UHECRs up the jet's Hillas limit are \emph{espresso}-accelerated, we can estimate their propagation time $t_{\rm prop}$ to be close to ballistic such that $t_{\rm prop} \sim \frac{H_{\rm jet}}{c}$. 
The effective optical depth $f_{p\gamma}$ can then be estimated as:
\begin{equation}
  \begin{split}
    f_{p\gamma}(E) \approx \langle t^{-1}_{p \gamma}(E) \rangle t_{\rm prop} \sim H_{\rm jet} \langle  n_\gamma (E) \rangle \sigma_{\rm eff} \xi 
   \end{split}
\end{equation}
where
\begin{equation}
    \langle  n_\gamma (E) \rangle \sim \frac{\epsilon_{\rm typ}L_{\epsilon_{\rm typ}}}{4\pi c}  \frac{1}{\epsilon_{\rm typ}} \frac{1}{H_{\rm jet}^2},
\end{equation}
where $\epsilon_{\rm typ}$ is the most likely photon energy, and $\epsilon_{\rm typ}$ depends on the considered particle energy such that $\epsilon_{\rm typ} \sim 0.5\bar{\epsilon}_{\rm res}/\gamma $ where $\bar{\epsilon}_{\rm res} \sim 0.34$GeV is the resonance energy and $\gamma=E/m_Ac^2$ is the Lorentz factor of a particle with energy $E$. Note that $\langle  n_\gamma (E) \rangle$ is energy dependent. For example, $\epsilon L_{\epsilon}\propto \epsilon^{2-\beta}$ (with $\beta\gtrsim1$), we have $\langle  n_\gamma (E) \rangle \propto E^{\beta-1}$.  

Finally we get an estimate for $E_\nu^2 F_{\rm UHE \nu}(E_\nu)$ such that:
\begin{equation}
    E_\nu^2F_{\rm UHE \nu}(E_\nu) \sim \frac{3}{8} \sigma_{\rm eff} \xi \frac{L_{\epsilon_{\rm typ}}(E)}{4\pi H_{\rm jet} c} E^2 F_{\rm UHECR}(E)
\end{equation}
In the example in the left panel of Figure~\ref{A_dependence}, $\lbol = 10^{48}\ergs$, $H_{\rm jet} \sim 1$kpc, and $\gamma > 10^7$ so $\epsilon_{\rm m} \sim 15$eV in the black hole frame; the values of $\alpha$, $\sigma_{\rm eff} $, and $\xi$ are discussed more in detail in Appendix~\ref{photomes_appendix}.

\section{Cosmological Evolution of AGN}\label{cosmo_appendix}

Following \cite{ueda+03,inoue+09b,harding+12} and \citetalias{murase+14}, We set the the cosmological evolution factor $\zeta_z(\lx, z)$, where $z$ the redshift and $\lx$ is the X-ray luminosity of the source:
\begin{equation}\label{eq:fz}
    \zeta_z(\lx, z)= 
\begin{cases}
    (1+z)^{4.23}& \text{if } z \leq z_l(\lx)\\
    [1+z_l(\lx)]^{4.23} \left[ \frac{1+z}{1+z_l(\lx)} \right]^{-1.5},& \text{if } z > z_l(\lx)
\end{cases}
\end{equation}
with 
\[
    z_l(\lx)= 
\begin{cases}
    z_c& \text{if } L_a \leq \lx\\
    z_c(\lx/L_a)^{0.335}& \text{if } L_a > \lx
\end{cases}
\]
such that $z_c = 1.9$ and $L_a = 10^{44.6}\ergs$. 

\bibliography{Total}
\bibliographystyle{aasjournal.bst}

\end{document}